\documentclass[11pt]{article}
\usepackage[dvips]{color}
\usepackage{epsfig}
\usepackage{amssymb, amsmath, multicol, multirow}

\renewcommand{\baselinestretch}{1.35}

\makeatletter

\def\singlespace{\def\baselinestretch{1}\@normalsize}

\newtheorem{theorem}{Theorem}[section]

\@addtoreset{equation}{section}
\renewcommand{\theequation}{\thesection.\arabic{equation}}

\renewcommand{\hat}{\widehat}

\makeatother
\def\c{\centerline}

\newcommand{\bx}{\mbox{\bf x}}

\newcommand{\bmu}{\mbox{\boldmath$\mu$}}
\newcommand{\bSigma}{\mbox{\boldmath$\Sigma$}}
\newcommand{\bpi}{\mbox{\boldmath$\pi$}}

\newcommand{\btheta}{\mbox{\boldmath$\theta$}}
\newcommand{\hbtheta}{\hat{\btheta}}
\newcommand{\hpi}{\hat{\pi}}

\newcommand{\etal}{{\it et al.}}

\newcommand{\bbeta}{\mbox{\boldmath$\beta$}}

\def\E{\mbox{E}}

\def\and{\mbox{and}}

\def\today{\ifcase\month\or
  January\or February\or March\or April\or May\or June\or
  July\or August\or September\or October\or November\or December\fi
  \space\number\day, \number\year}

\newcommand{\beq}{\begin{equation}}
\newcommand{\eeq}{\end{equation}}
\newcommand{\beqnn}{\begin{eqnarray*}}
\newcommand{\eeqnn}{\end{eqnarray*}}
\newcommand{\beqn}{\begin{eqnarray}}
\newcommand{\eeqn}{\end{eqnarray}}

\begin{document}

\title{Model Selection for Gaussian Mixture Models}

\author{Tao Huang, \ \ Heng Peng \ \ and Kun Zhang  \footnotemark}

\maketitle
\begin{singlespace}
\begin{footnotetext}{
Tao Huang is Assistant Professor, Department of Statistics,
University of Virginia, Charlottesville, VA 22904. Heng Peng is
Assistant Professor, Department of Mathematics, Hong Kong Baptist
University, Kowloon Tong, Hong Kong.  Kun Zhang is  Research
Scientist, Department of Sch\"{o}lkopf,   Max Planck Institute for
Biological Cybernetics, Spemannstrasse 38, 72076, T\"{u}bingen.
}
\end{footnotetext}
\end{singlespace}

\begin{abstract}
This paper is concerned with an important issue in finite mixture
modelling, the selection of the number of mixing components. We
propose a new penalized likelihood method for model selection of
finite multivariate  Gaussian mixture models.  The proposed method
is shown to be statistically consistent in determining of the number
of components. A modified EM algorithm is developed to
simultaneously select the number of components and to estimate the
mixing weights, i.e. the mixing probabilities, and unknown
parameters of Gaussian distributions. Simulations and a real data
analysis are presented to illustrate the performance of the proposed
method.
\end{abstract}


\section{Introduction}

Finite mixture modeling is a flexible and powerful approach to
modeling data that is heterogeneous and stems from multiple
populations, such as data from patter recognition, computer vision,
image analysis, and machine learning. The Gaussian mixture model is
an important mixture model family. It is well known that any
continuous distribution can be approximated arbitrarily well by a
finite mixture of normal densities  (Lindsay, 1995; McLachlan and
Peel, 2000). However, as demonstrated by Chen (1995), when the
number of components is unknown, the optimal convergence rate of the
estimate of a finite mixture model is slower than the optimal
convergence rate when the number is known. In practice, with too
many components, the mixture may overfit the data and yield poor
interpretations, while with too few components, the mixture may not
be flexible enough to approximate the true underlying data
structure. Hence, an important issue in finite mixture modeling is
the selection of the number of components, which is not only of
theoretical interest, but also significantly useful in practical
applications.

Most conventional methods for determining the order of the finite
mixture model are based on the likelihood function and some
information theoretic criteria, such as AIC and BIC. Leroux (1992)
investigated the properties of AIC and BIC for selecting the number
of components for finite mixture models and showed that these
criteria would not underestimate the true number of components.
Roeder and Wasserman (1997) showed the consistency of BIC when a
normal mixture model is used to estimate a density function
``nonparametrically". Using the locally conic parameterization
method developed by Dacunha-Castelle and Gassiat (1997), Keribin
(2000) investigated the consistency of the maximum penalized
likelihood estimator for an appropriate penalization sequence.
Another class of methods is based on the distance measured between
the fitted model and the nonparametric estimate of the population
distribution, such as penalized minimum-distance method (Chen and
Kalbfleisch, 1996), the Kullback-Leibler distance method (James,
Priebe and Marchette, 2001) and the Hellinger distance method (Woo
and Sriram, 2006). To avoid the irregularity of the likelihood
function for the finite mixture model that emerges when the number
of components is unknown, Ray and Lindsay (2008) suggested to use a
quadratic-risk based approach to select the number of components for
the finite multivariate mixture. However, these methods are all
based on the complete model search algorithm and the computation
burden is heavy. To improve the computational efficiency, recently,
Chen and Khalili (2008) proposed a penalized likelihood method with
the SCAD penalty (Fan and Li, 2001) for mixtures of univariate
location distributions. They proposed to use the SCAD penalty
function to penalize the differences of location parameters, which
is  able to merge some subpopulations by shrinking such differences
to zero. However, similar to most conventional order selection
methods, their penalized likelihood method can be only used for  one
dimensional location mixture models. Furthermore, if some components
in the true/optimal model have the same location (which is the case
for the experiment in Subsection 4.2 of this study), some of them
would be eliminated incorrectly by this method.


On the other hand, Bayesian approaches have been also used to find a
suitable number of components of the finite mixture model. For
instance, variational inference, as an approximation scheme of
Bayesian inference, can be used to determine the number of the
components in a fully Bayesian way (see, e.g., Corduneanu and C.M.
Bishop (2001) or Chapter 10.2 of Bishop (2006)). Moreover, with
suitable priors on the parameters, the maximum a posteriori (MAP)
estimator can be used for model selection. In particular, Ormoneit
and Tresp (1998) and Zivkovic and van der Heijden (2004) put the
Dirichlet prior on the mixing weights, i.e. the mixing
probabilities, of the components in the Gaussian mixture model, and
Brand (1999) applied the ``entropic prior" on the same parameters to
favor models with small entropy. They then used the MAP estimator to
drive the mixing weights associated with unnecessary components
toward extinction. Based on an improper Dirichlet prior, Figueiredo
and Jain (2002) suggested to use minimum message length criterion to
determine the number of the components, and further proposed an
efficient algorithm for learning a finite mixture from multivariate
data. We would like to point out the significant difference between
those approaches and our proposed method in this paper. When a
component is eliminated, our suggested objective function changes
continuously, while those approaches encounter a sudden change in
the objective function because $zero$ is not in the support area of
the prior distribution for the mixing weights, such as the Dirichlet
prior. Therefore, it is difficult to study statistical properties of
these Bayesian approaches and, especially, the consistency analysis
is often missing in the literature.

In this paper, we propose a new penalized likelihood method for
finite mixture models. In particular, we focus on finite Gaussian
mixture models.   Intuitively, if some of the mixing weights or
mixing probabilities are shrunk to zero, the corresponding
components are eliminated and a suitable number of components is
retained.  By doing this, we can deal with multivariate Gaussian
mixture models and do not need to assume common covariance matrix
for different components. Popular $L_p$ types of penalty functions
would suggest to penalize the mixing weights directly. However, we
will show that such types of penalty functions do not penalize the
mixing weights severely enough and cannot shrink them to zero.
Instead, we propose to penalize the logarithm of mixing weights.
When some mixing weights are shrunk to zero, the objective function
of the proposed method changes continuously, and hence we can
investigate its statistical properties, especially the consistency
of the proposed penalized likelihood method.

The rest of the paper is organized as follows. In Section 2, we
propose a new penalized likelihood method for finite multivariate
Gaussian mixture models. In Section 3, we derive asymptotic
properties of the estimated number of components.  In Section 4,
simulation studies and a real data analysis are presented to
illustrate the performance of our proposed methods.  Some
discussions are given in Section 5. Proof will be delegated in the
Appendix.

\section{Gaussian Mixture Model Selection}

\subsection{Penalized Likelihood Method}

Gaussian mixture model (GMM) models the density of a $d$-dimensional random variable  $\bx$ as a weighted sum of some Gaussian densities
\begin{equation} \label{e2.1}
f(\bx) = \sum_{m=1}^{M}  \pi_m \phi(\bx; \bmu_m, \bSigma_m),
\end{equation}
where $\phi(\bx; \bmu_m, \bSigma_m)$ is a Gaussian density with mean
vector $\bmu_m$ and covariance matrix $\bSigma_m$, and $\pi_m$ are
the positive mixing weights or mixing probabilities that satisfy the
constraint $\sum_{m=1}^{M} \pi_m=1$. For identifiability of the
component number, let $M$ be the smallest integer such that all
components are different and the mixing weights are nonzero. That
is, $M$ is the smallest integer such that  $\pi_m >0$ for $1 \leq m
\leq M$, and $(\bmu_a, \bSigma_a) \neq (\bmu_b, \bSigma_b)$ for
$1\leq a \neq b \leq M$. Given the number of components $M$, the
complete set of parameters of GMM, $\btheta = \{ \bmu_1, \bSigma_1,
\cdots, \bmu_M, \bSigma_M, \pi_1, \cdots, \pi_M\}$, can be
conveniently estimated by maximum likelihood method via the EM
algorithm. To avoid overfitting and underfitting, an important issue
is to determine the number of components $M$.

Intuitively, if some of the mixing weights are shrunk to zero, the
corresponding components are eliminated and a suitable number of
components is retained.  However, this can not be achieved by
directly penalizing mixing weights $\pi_m$.  By considering the
indicator variables $y_{im}$ that show if the $i$th observation
arises from the $m$th component as missing data, one can find the
expected complete-data log-likelihood function (pp. 48, McLachlan
and Peel, 2000):
\begin{eqnarray} \label{e2.2}
\E(\ell(\btheta)) &=& \E\left\{ \log \prod_{i=1}^{n} f(\bx_i; \btheta) \right\} = \E \left\{\sum_{i=1}^{n} \sum_{m=1}^{M}  y_{im} \left[ \log \pi_m + \log \phi(\bx_i; \bmu_m, \bSigma_m) \right] \right\} \nonumber \\
&=& \sum_{i=1}^{n} \sum_{m=1}^{M} h_{im} \log \pi_m + \sum_{i=1}^{n} \sum_{m=1}^{M} h_{im} \log \phi(\bx_i; \bmu_m, \bSigma_m),
\end{eqnarray}
where $\ell(\btheta)$ is the complete-data log-likelihood, and
$h_{im}$ is the posterior probability that the $i$th observation
belongs to the $m$th component. Note that the expected complete-data
log-likelihood involves $\log \pi_m$, whose gradient grows very fast
when $\pi_m$ is close to zero.  Hence the popular $L_p$ types of
penalties may not able to set insignificant $\pi_m$ to zero.

\begin{figure}[htp]
\vspace{-1in}
\begin{tabular}{cc}
\hspace{-1in} \includegraphics[height=4in,
width=5in]{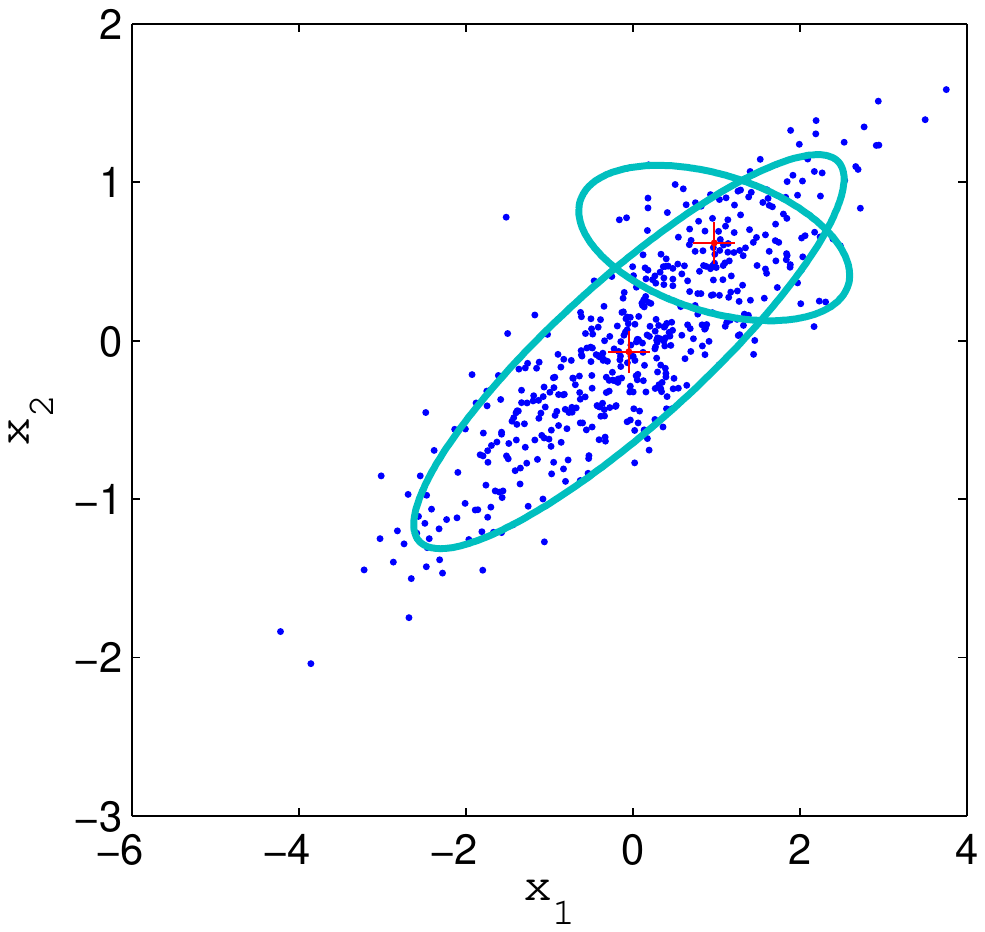} & \hspace{-1.75in}
\includegraphics[height=4in, width=5in]{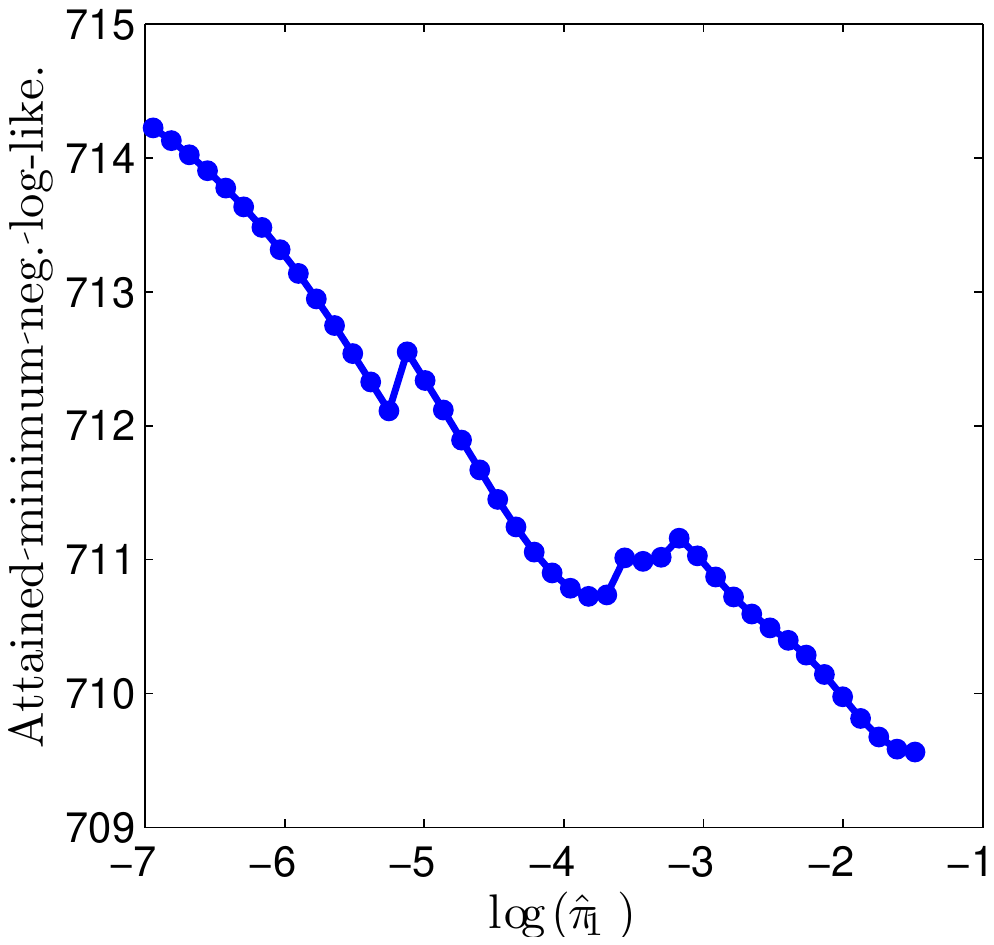}  \vspace{-1.25in} \\
\hspace{-1in}(a)& \hspace{-1.75in}(b) \\
\end{tabular}
\begin{singlespace}
\caption{\em \label{gmm2c} An illustration on the behavior of the negative log-likelihood function when a mixing probability is close to zero. (a) The simulated data set and a learned two-component GMM model. (b) The minimized negative log-likelihood as a function of $\log{{\pi}_1}$.  Note that the $x$-axis is in $\log$ scale.}
\end{singlespace}
\end{figure}

Below we give a simple illustration on how the likelihood function
changes when a mixing probability approaches to zero.  In
particular,  a data set of 1000 points is randomly generated from a
bivariate Gaussian distribution (i.e., a GMM with only one
component). A GMM with two components, $f(\mathbf{x}) = \pi_1
\phi(\mathbf{x}; \boldsymbol{u}_1, \boldsymbol{\sigma}_1) +
(1-\pi_1) \phi(\mathbf{x}; \boldsymbol{u}_2,
\boldsymbol{\Sigma}_2)$,  is then used to fit the data.   The
learned two Gaussian components are depicted in
Figure~\ref{gmm2c}(a), and $\hat{\pi}_1$ is 0.227. Furthermore, to
see how the negative likelihood function changes with respective to
it, let ${\pi}_1$ gradually approach zero.  For each fixed
${\pi}_1$, we optimize all other parameters, $\{\boldsymbol{\mu}_i,
\boldsymbol{\Sigma}_i, i=1,2\}$, by maximizing the likelihood
function. Figure~\ref{gmm2c}(b) depicts how the minimized negative
log-likelihood function changes with respective to $\log{\pi_1}$. It
shows that the log-likelihood function changes almost linearly along
with $\log({\pi}_1)$ when $\pi_1$ is close to zero, albeit some
small upticks.  In other words, the derivative of the log-likelihood
function with respective to $\hat{\pi}_1$ is approximately
proportional to $1/{\hat{\pi}_1}$ when $\hat{\pi}_1$ is close to
zero, and it would dominate the derivative of $\pi_1^p$.
Consequently $L_p$ penalties can not set insignificant $\pi_1$ to
zero.

By the discussion above,  we know that $L_1$-type penalized
likelihood methods are not omnipotent, especially when the model is
not a regular statistical model.  In fact, the expected
complete-data log-likelihood function (\ref{e2.2}) suggests that we
need to consider some types of penalties on $\log \pi_m$ in order to
achieve the sparsity for $\bpi=\{\pi_1, \cdots, \pi_M\}$. In
particular, we simply choose to penalize $\log(\frac{\epsilon +
\pi_m}{\epsilon}) = \log(\epsilon + \pi_m) - \log(\epsilon)$, where
$\epsilon$ is a very small positive number, say $10^{-6}$ or
$o(n^{-\frac12}\log^{-1} n)$ as the discussion of Theorem 3.2. Note
that $\log(\epsilon + \pi) - \log(\epsilon)$ is a monotonically
increasing function of $\pi$, and it is shrunk to zero when the
mixing weight $\pi$ goes to zero. Therefore, we propose the
following penalized log-likelihood function,
\begin{equation} \label{e2.3}
\ell_P (\btheta) = \ell(\btheta) - n \lambda D_f \sum_{m=1}^{M} \left[\log(\epsilon + \pi_m) - \log(\epsilon) \right],
\end{equation}
where $\ell(\btheta)$ is the log-likelihood function, $\lambda$ is a
tuning parameter, and $D_f$ is the number of free parameters for
each component.  For GMM with arbitrary covariance matrices, each
component has $D_f = 1 + d + d(d+1)/2 = d^2/2 + 3d/2 + 1$ number of
free parameters. Although here $D_f$ is a constant and can be
removed from the equation above, it would simplify the search range
of $\lambda$ in the numerical study.

Note our penalty function is similar to that derived with the
Dirichlet prior from Bayesian point of view, both using logarithm
function of the mixing weights of the finite mixture model as the
penalty function or prior distribution function. However, for the
Dirichlet prior, the objective penalized likelihood function
penalizes $\log \pi_i$, and unlike our proposed penalty function
$\log(\epsilon + \pi_m) - \log(\epsilon)$, zero is not in the
support area of the penalty function $\log \pi_i$. In the
mathematical sense, these Bayesian approaches can not shrink the
mixing weights to zero exactly since zero is not well defined for
the objective function. In other words, the objective function is
not continuous when some of mixing weights shrunk continuously to
zero. As the discussion by Fan and Li (2001), such discontinuity
poses challenges to investigate the statistical properties of
related penalized or Bayesian methods. It is be one of main reasons
why there is little literature on studies the consistency of the
proposed methods based on the Dirichlet prior. Moveover, when
$\epsilon=0$, our penalty function can not be seen as in a border
family of Dirichlet distributions. In fact, when $\epsilon=0$, there
is no definition for the second term of our proposed penalty
function. Though when $\epsilon>0$, our proposed penalty function is
similar to the one used by Figueirdo and Jain (2002). However, zero
is still not in the support area of their proposed improper prior
function, and their method has similar problems as those Bayesian
approaches based on the Dirichlet prior do.

As discussed in Fan and Li (2001), a
good penalty function should yield an estimator with three
properties: unbiasedness, sparsity, and continuity. It is obvious
that $\log(\frac{\epsilon + \pi_m}{\epsilon})$ would over penalize large $\pi_m$ and yield a biased estimator.
Hence, we also consider the following penalized log-likelihood function,
\begin{equation} \label{e2.4}
\ell_P (\btheta) = \ell(\btheta) - n \lambda D_f \sum_{m=1}^{M}
\left[\log(\epsilon + p_\lambda(\pi_m)) - \log(\epsilon) \right].
\end{equation}
Compared to (\ref{e2.3}), the only difference is that
$\pi_m$ is replaced by  $p_\lambda(\pi_m)$ in the penalty function, where
$p_\lambda(\pi)$ is the SCAD penalty function proposed by Fan and Li
(2001) and is conveniently characterized through its
derivative:
$$
p'_\lambda(\pi)=I(\pi \le
\lambda)+\frac{(a\lambda-\pi)_+}{(a-1)\lambda} I (\pi>\lambda),
$$
for some $a>2$ and $\pi>0$.  It is easy to see that, for a
relatively large $\pi_m$ and $\pi_m>a\lambda$, $p_\lambda(\pi_m)$ is
a constant, and henceforth the estimator of this $\pi_m$ is expected
be unbiased.

\subsection{Modified EM Algorithm}

Here we propose a modified EM algorithm to maximize (\ref{e2.3}) and
(\ref{e2.4}) iteratively in two steps.

First we introduce a modified EM algorithm to maximize (\ref{e2.3}).
By (\ref{e2.2}) and (\ref{e2.3}), the expected penalized
log-likelihood function is
\begin{equation} \label{e2.5}
\sum_{i=1}^{n} \sum_{m=1}^{M} h_{im} \log \pi_m + \sum_{i=1}^{n} \sum_{m=1}^{M} h_{im} \log \phi(\bx_i; \bmu_m, \bSigma_m) - n \lambda D_f \sum_{m=1}^{M} \left[ \log(\epsilon + \pi_m) - \log(\epsilon) \right].
\end{equation}

In the E step, we calculate the posterior probability given the current estimate $\hbtheta=(\hat \bmu_1, \hat \bSigma_1, \ldots, $ $\hat \bmu_M, \hat \bSigma_M, \hat \pi_1, \ldots, \hat \pi_M)$
\begin{equation*}
h_{im} = \frac{\hat \pi_m \phi(\bx_i; \hat \bmu_m, \hat \bSigma_m)}{\sum_{m=1}^{M} \hat \pi_m \phi(\bx_i; \hat \bmu_m, \hat \bSigma_m)}.
\end{equation*}
In the M step, we update  $\btheta = \{ \bmu_1, \bSigma_1,  \cdots,
\bmu_M, \bSigma_M, \pi_1, \cdots, \pi_M\}$ by maximizing the
expected penalized log-likelihood function (\ref{e2.5}).  Note that
we can update $\{\pi_1, \cdots, \pi_M\}$ and  $\{\bmu_1, \bSigma_1,
\cdots, \bmu_M, \bSigma_M\}$ separately as they are not intervened
in (\ref{e2.5}).  To obtain an estimate for $\bpi=(\pi_1, \cdots,
\pi_M)$, we introduce a Lagrange multiplier $\beta$ to take into
account for the constraint $\sum_{m=1}^{M} \pi_m= 1$, and aim to solve
the following set of equations,
\begin{equation*}
\frac{\partial}{\partial \pi_m} \left[\sum_{i=1}^{n} \sum_{m=1}^{M} h_{im} \log \pi_m - n \lambda D_f \sum_{m=1}^{M}  \log(\epsilon + \pi_m) - \beta (\sum_{m=1}^{M} \pi_m - 1) \right] = 0.
\end{equation*}
Given $\epsilon$ is very close to zero, by straightforward calculations, we obtain
\begin{equation} \label{e2.6}
\hpi_m = \max \left\{0, \frac{1}{1- M \lambda D_f} \left[ \frac{1}{n} \sum_{i=1}^{n} h_{im} - \lambda D_f \right] \right\}.
\end{equation}
The update equations on $\{\bmu_1, \bSigma_1, \cdots, \bmu_M, \bSigma_M\}$  are the same as those of the standard EM algorithm for GMM (McLachlan and Peel, 2000).  Specifically, we update $\bmu_m$ and $\bSigma_m$ as follows,
\begin{equation*}
\hat \bmu_m = \sum_{i=1}^{n} h_{im} \bx_i / \sum_{i=1}^{n} h_{im}, \qquad
\hat \bSigma_m =   \sum_{i=1}^{n} h_{im} (\bx_i - \hat \bmu_m)(\bx_i - \hat \bmu_m)^T / \sum_{i=1}^{n} h_{im}.
\end{equation*}

In summary the proposed modified EM algorithm works as follows: it
starts with a pre-specified large number of components, and whenever
a mixing probability is shrunk to zero by (\ref{e2.6}), the
corresponding component is deleted, thus fewer components
are retained for the remaining EM iterations.  Here we abuse the
notation $M$ for the number of components at beginning of each EM
iteration, and through the updating process, $M$ becomes smaller and
smaller. For a given EM iteration step, it is possible that none, one, or
more than one components are deleted.

The modified EM algorithm for maximizing (\ref{e2.4}) is similar to the one for (\ref{e2.3}), and the only
difference is in the M step for maximizing $\bpi$.
Given the current estimate $(\pi_1^0,\ldots,\pi_M^0)$ for $\bpi$, to solve
\begin{equation*} \frac{\partial}{\partial \pi_m}
\left[\sum_{i=1}^{n} \sum_{m=1}^{M} h_{im} \log \pi_m - n \lambda
D_f \sum_{m=1}^{M}  \log(\epsilon + p_\lambda(\pi_m)) - \beta
(\sum_{m=1}^{M} \pi_m - 1) \right] = 0,
\end{equation*}
we substitute $\log(\epsilon + p_\lambda(\pi_m))$ by its linear approximation $\log(\epsilon +
p_\lambda(\pi^0_m))+\frac{p_\lambda'(\pi_m^0)}{\epsilon+p_\lambda(\pi^0_m)}
(\pi_m-\pi_m^0)$ in the equation above. Then by straightforward calculations, $\pi_m$ can be
updated as follows
\begin{equation}\label{PI}
\pi_m=\frac{1}{T_m}\sum\limits_{i=1}^n h_{mi}
\end{equation}
where
$$ T_m=n-n\lambda D_f \sum\limits_{m=1}^M
\frac{p_\lambda'(\pi_m^0)\pi_m^0}{\epsilon+p_\lambda(\pi^0_m)}+n
\lambda D_f \frac{p_\lambda'(\pi_m^0)}{\epsilon+p_\lambda(\pi^0_m)}.
$$
If an updated $\pi_m$ is smaller than a pre-specified small
threshold value, we then set it to zero and remove the corresponding
component from the mixture model. In
 numerical study, (\ref{PI}) is seldom
exactly equal to zero. Using such threshold method to set
 mixing weight to zero is only to avoid the numerical unstabilities.
 Because of the consistent properties of
such penalty function derived in Section 3, we can set this
pre-specified small threshold value as small as possible, though
small threshold value would increase iterative steps of EM algorithm
and computation time. In our numerical studies, we set this
threshold as $10^{-4}$, but the smallest mixing weight in the
following two numerical examples are $1/600$ and $1/1000$, both
larger than this threshold.

\subsection{Selection of Tuning Parameters}

To obtain the final estimate of the mixture model by maximizing
(\ref{e2.3}) or (\ref{e2.4}), one needs to select the tuning
parameters $\lambda$ and $a$ (the latter is involved (\ref{e2.4})).
Our simulation studies show that the numerical results are not
sensitive to the selection of $a$ and therefore by the suggestion of
Fan and Li (2001) we set $a=3.7$.  For standard LASSO and SCAD
penalized regressions, there are many methods to select $\lambda$,
such as generalized cross-validation (GCV) and BIC (See Fan and Li,
2001 and Wang \etal, 2007). Here we define a BIC value
$$
\mathrm{BIC}(\lambda)=\sum\limits_{i=1}^n
\log\left\{\sum_{m=1}^{\hat{M}} \pi_m \phi(\bx_i; \hat{\bmu}_m,
\hat{\bSigma}_m)\right\}-\frac12 \hat{M} D_f \log n
$$
and select $\hat{\lambda}$ by
$$
\hat{\lambda}=\mbox{arg}\max\limits_{\lambda} \mbox{BIC}(\lambda),
$$
where $\hat{M}$ is
the estimate of the number of components and $\hat{\bmu}_m$ and
$\hat{\bSigma}_m$ are the estimates of $\bmu_m$ and $\bSigma_m$
 for maximizing $(\ref{e2.3})$ or $(\ref{e2.4})$ for a given  $\lambda$.

\section{Asymptotic Properties}
It is possible to extend our proposed model selection method to more
generalized mixture models, However, to illustrate the basic idea of
the proposed method without many mathematical difficulties, in this
section, we only show the model selection consistency of the
proposed method for Gaussian mixture models.

First, we assume that, for the true Gaussian mixture model, there
are $q$ mixture components, $q \le M$ with $\pi_i=0$, for
$i=1,\ldots,M-q$, $\pi_i=\pi^0_l$, for $i=M-q+1,\ldots,M,
l=1,\ldots,q$. Then by the idea of locally conic models
(Dacunha-Castelle and Gassiat, 1997 and 1999), the density function
of the Gaussian mixture model can be rewritten as
$$
f(\bx,\btheta)=f(\bx, \theta,\bbeta)= \sum\limits^{M-q}_{i=1}
\lambda_i \theta \cdot \phi(\bmu_i,\bSigma_i)+\sum\limits_{l=1}^q
(\pi_l^0+\rho_l \theta) \cdot \phi(\bmu^0_l+\theta
\delta_\mu^l,\bSigma_l^0+\theta \delta_{\Sigma}^l).
$$
where
$$
\bbeta=(\lambda_1,\ldots,\lambda_{M-q},
\bmu_1,\ldots,\bmu_{M-q},\bSigma_1,\ldots,\bSigma_{M-q},
\delta_{\mu}^1,\ldots, \delta_{\mu}^q,
\delta_{\Sigma}^1,\ldots,\delta_{\Sigma}^q, \rho_1,\ldots,\rho_q),
$$
$\bmu^0_i,i=1,\ldots,q,$ and $\bSigma_i^0,i=1,\ldots,q,$ are the
true values of multivariate normal components, and
$(\pi_1,\ldots,\pi_M)$ in the original Gaussian mixture model can be
defined as $\pi_i=\lambda_i\theta, i=1,\ldots,M-q$ and
$\pi_i=\pi_l^0+\rho_l\theta, i=M-q+1,\ldots,M, l=1,\ldots,q$.
Similar to Dacunha-Castelle and Gassiat (1997, 1999), by the
restrictions imposed on the $\bbeta$:
\begin{eqnarray*}
\lambda_i \ge 0, \ \mu_i \in \mathbf{R}^d, &\mbox{and}&  \bSigma_i
\in
\mathbf{R}^{d \times d}, i=1,\ldots, M-q, \\
\delta_{\mu}^l \in \mathbf{R}^d, \ \delta_{\Sigma}^l \in
\mathbf{R}^{d\times d}, & \mbox{and} & \rho \in \mathbf{R}, \
l=1,\ldots, q, \\
\sum\limits_{i=1}^{M-q} \lambda_i+\sum\limits_{l=1}^q \rho_l=0
&\mbox{and}&\sum\limits_{i=1}^{M-q}\lambda_i^2+\sum\limits_{l=1}^q
\rho_l^2+\sum\limits_{l=1}^q
\|\delta^l_{\mu}\|^2+\sum\limits_{l=1}^q \|\delta^l_{\Sigma}\|^2 =1,
\end{eqnarray*}
and by the permutation, such a parametrization is locally conic and
identifiable.

After the parametrization, the penalized likelihood functions
(\ref{e2.3}) and  (\ref{e2.4}) can be written as
\begin{eqnarray} \nonumber
\ell_P (\btheta) &=& \ell(\btheta) - n \lambda D_f \sum_{m=1}^{M}
\left[\log(\epsilon + \pi_m)) - \log(\epsilon) \right] \\
\nonumber
&\hat{=}& \ell_P(\theta,\bbeta) \\
 &=& \sum\limits_{i=1}^n \log
f(\bx_i, \theta,\bbeta) - n \lambda D_f \sum_{m=1}^{M}
\left[\log(\epsilon + \pi_m) - \log(\epsilon) \right],
\end{eqnarray}
and
\begin{eqnarray} \nonumber \ell_P (\btheta) &=& \ell(\btheta) - n
\lambda D_f \sum_{m=1}^{M}
\left[\log(\epsilon + p_\lambda(\pi_m)) - \log(\epsilon) \right] \\
\nonumber
&\hat{=}& \ell_P(\theta,\bbeta) \\
 &=& \sum\limits_{i=1}^n \log
f(\bx_i, \theta,\bbeta) - n \lambda D_f \sum_{m=1}^{M}
\left[\log(\epsilon + p_\lambda(\pi_m)) - \log(\epsilon) \right],
\end{eqnarray}
respectively.

We need the following conditions to derive the asymptotic
properties of our proposed method.
\begin{itemize}
\item[P1:] $\|\mu_i\| \le C_1,  \|\bSigma_i\| \le C_2,
i =1,\ldots,M,$ where $C_1$ and $C_2$ are large enough constants.
\item[P2:] $\min\limits_{i,k} \{\lambda_k(\bSigma_i),k=1,\ldots,d, i=1,\ldots, M\} \ge
C_3$, where $\lambda_k(\bSigma_i)$ are the eigenvalues of
$\bSigma_i$ and $C_3$ is a positive constant.
\end{itemize}

Compared to the conditions in Dacunha-Castelle and Gassiat (1997,
1999), the conditions P1 and P2 are slightly stronger. Without lose
of generality, we assume that the parameters in the mixture model
are in a bounded compact space not only for mathematical
conveniences, but also for avoiding the identifiability and
ill-posedness problems of the finite mixture model as discussed in
Bishop (2006). Those conditions are also practically reasonable for
our revised EM algorithm as the discussion in Figueirdo and Jain
(2002).

First, even if it is known there is $K$ mixture components in the
model, the maximum likelihood or penalized maximum likelihood
solution still has a total of $K!$ equivalent solutions
corresponding to the $K!$ ways of assigning $K$ sets of parameters
to $K$ components. Although this is an important issue when we wish
to interpret the estimate parameter values for a selected model.
However, the main focus of our paper is to determine the order and
find a good density estimate with the finite mixture model. The
identifiability problem is irrelevant as all equivalent solutions
have same estimate of the order and the density function.

Second, Condition (P2) is imposed to guarantee the non-singularity
of the covariance matrices, avoiding the ill-posedness in the
estimation of the finite multivariate Gaussian mixture model with
unknown covariance matrices using our proposed penalized maximum
likelihood method. Similar as the discussion in Figueirdo and Jain
(2002), given a sufficient large number of the initial mixture
components, say M, our proposed modified EM algorithm selects
components in a backwards manner by merging smaller components into
a large component, and thus reducing the number of components. On
the other hand, pre-specifying an extreme large value for M should
be avoided as well. In our numerical studies, the initial number of
mixture components M is set to be smaller than n/p so that each
estimated mixture component has a positive covariance matrix, as
required by Condition (P2).

\begin{theorem} Under the conditions (P1) and (P2) and if $\sqrt{n}\lambda \to \infty$,  $\lambda \to 0$ and $\epsilon=o(1/\sqrt{n})$,   there exists a
local maximizer $(\theta,\bbeta)$ of $\ell_P$, which was given in
(3.2), such that $\theta=O_p(1/\sqrt{n})$, and for such local
maximizer, the number of the mixture components $\hat{q}_n \to q $
with probability tending to one.
\end{theorem}

\begin{theorem} Under the conditions (P1) and (P2) and if $\sqrt{n}\lambda \to C$  and $\epsilon=o(\frac{1}{\sqrt{n}\log n})$ where $C$ is a constant,   there exists a
local maximizer $(\theta,\bbeta)$ of $\ell_P$, which was given in
(3.1), such that $\theta=O_p(1/\sqrt{n})$, and for such local
maximizer, the number of the mixture components $\hat{q}_n \to q $
with probability tending to one.
\end{theorem}

\noindent{\bf Remarks:} Under Conditions (P1) and (P2) and with an
appropriate tuning parameter $\lambda$,  the consistency of our
proposed two penalized methods have been shown by the two theorems
above. Although maximizing (3.2) using our proposed EM algorithm is
a bit complicated than that for (3.1), it has theoretical
advantages.  Unlike (3.2), a mixture component with a relative large
mixing weight $\pi_i$ is still penalized in (3.1) by the penalty
function $\log(\pi_i+\epsilon)-\log \epsilon$, and this would
produce a bias model estimation and affect the consistency of the
model selection as the discussion by Fan and Li (2001).  Moreover,
in practice it is easier to select an appropriate tuning parameter
for (3.2) than for (3.1) to guarantee the consistency of the final
model selection and estimation. In particular, the following theorem
shows that the proposed BIC criterion always selects the reasonable
tuning parameter with probability tending one by which maximizing
(3.2) selects the consistent component number of the finite Gaussian
mixture model.

Let $\mathrm{Component}_\lambda$ denote the number of component of
Gaussian mixture models selected by (3.2) with the tuning parameter
$\lambda$, and $\lambda_{BIC}$ is the lambda selected by the
proposed BIC criterion in Section 2.3. Then we have the following
theorem.

\begin{theorem}
Under the conditions (P1) and (P2),
$\mathrm{Pr}(\mathrm{Component}_{\lambda_{BIC}}=q) \to 1$.
\end{theorem}

The proofs of the theorems are given in the appendix.

\section{Numerical Studies}

\subsection{Example I}
In the first example, we generate 600 samples from a three-component bivariate normal mixture 
with mixing weights $\pi_1=\pi_2=\pi_3=1/3$, mean vectors
$\bmu_1=\left[-1, 1\right]^T$, $\bmu_2=\left[1, 1\right]^T$,
$\bmu_3=\left[0, -\sqrt{2}\right]^T$, and covariance matrices
$\bSigma_1=[0.65, ~0.7794;~0.7794,~ 1.55] $, $\bSigma_2=[0.65,
 -0.7794;~ -0.7794,~ 1.55]$, $\bSigma_3=\text{diag}\{2, 0.2\}$.
 In fact, these three components are obtained by rotating and shifting  a
 common Gaussian density $\mathcal{N}(\mbox{\boldmath{0}},\textrm{diag}(2, 0.2))$, and
 together they exhibit a triangle shape.

\begin{figure}[htbp]
\c{\psfig{figure=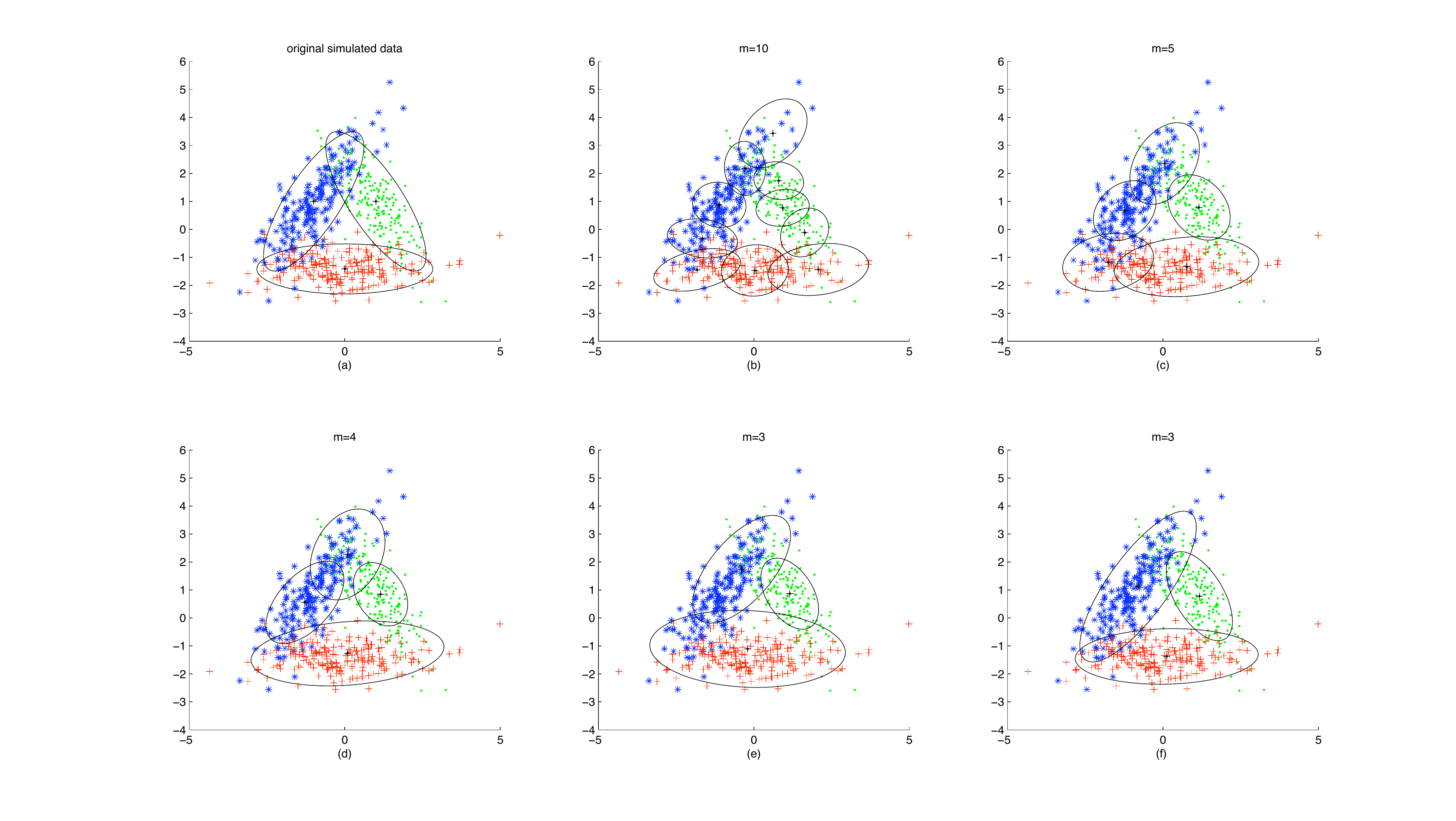,height=4in,width=6in}}
\begin{singlespace}
\caption{\em  \label{fig2s1} One typical run.   (a) a simulated data
set. (b) initialization for $m=10$ components, (c-e) three
intermediate estimates for $M=6, 5, 4$, respectively, (f) the final
estimate for $M=3$. }
\end{singlespace}
\end{figure}

\begin{figure}[htbp]
\c{\psfig{figure=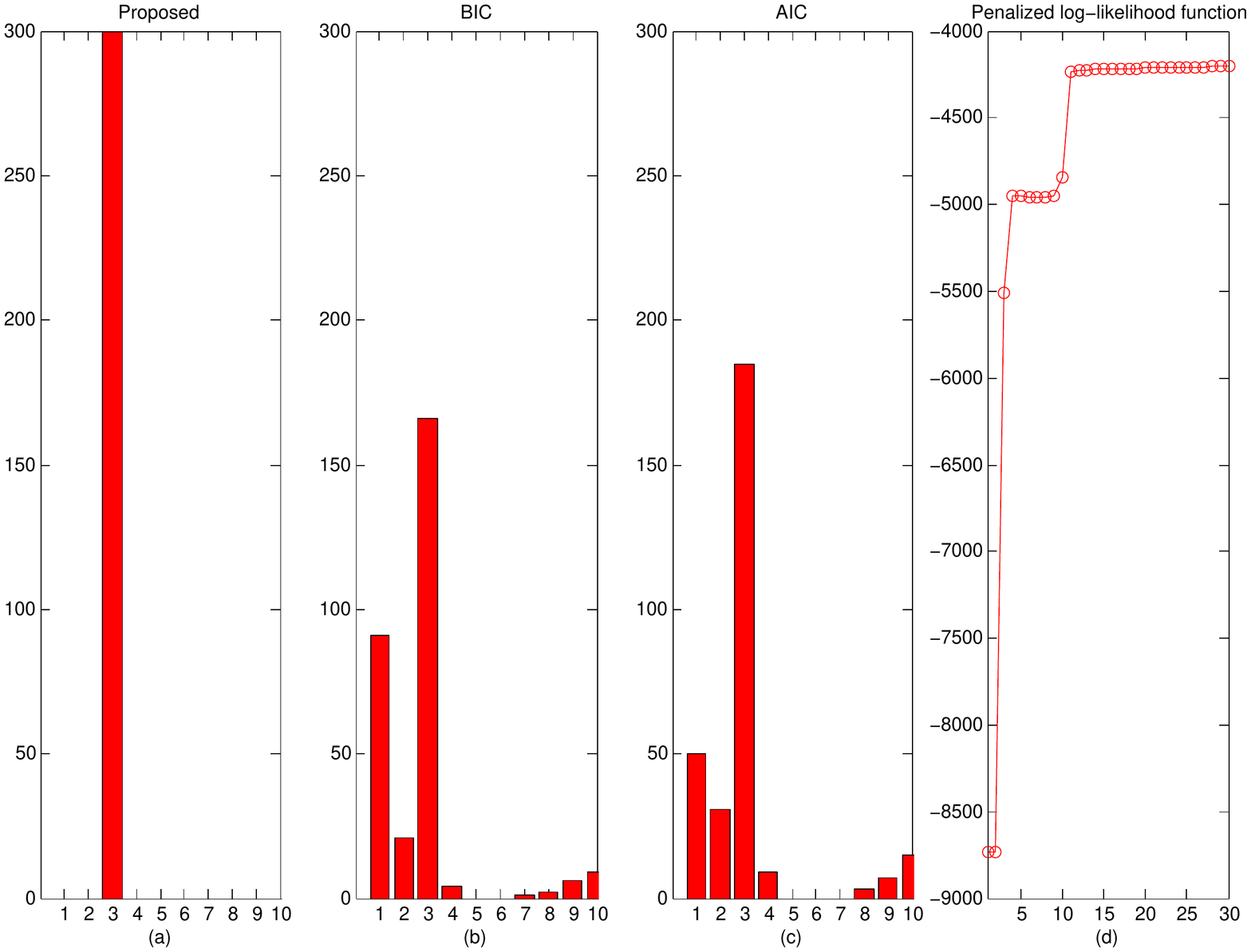,height=3in,width=6in}}
\begin{singlespace}
\caption{\em  \label{fig3s1} Histogram of  estimated numbers of
components. (a) the proposed method (2.3), (b) BIC, (c) AIC.  (d)
The penalized log likelihood function for one typical run. }
\end{singlespace}
\end{figure}

We run both of our proposed penalized likelihood methods
(\ref{e2.3}) and (\ref{e2.4}) for 300 times. The maximum initial
number of components is set to be 10 or 50 , the initial value for
the modified EM algorithms is estimated by K-means clustering, and
the tuning parameter $\lambda$ is selected by our proposed BIC
method. Figure~\ref{fig2s1} shows the evolution of the modified EM
algorithm, with the maximum number of components as 10.  We compare
our proposed methods with traditional AIC and BIC methods.
Figure~\ref{fig3s1}(a-c) shows the histograms of the estimated
component numbers.  One can see that our proposed methods perform
much better in identifying the correct number of components than AIC
and BIC methods do. In fact, both proposed methods  estimate the
number of components $100\%$ correctly regardless of the maximum
initial number of components. Figure~\ref{fig3s1}(d) depicts the
evolution of the penalized log-likelihood function (\ref{e2.3}) for
the simulated data set in Figure~\ref{fig2s1}(a) in one run, and
shows how our proposed modified EM algorithm converges numerically.

In addition,  when the number of components is correctly identified,
we summarize the estimation of the unknown parameters of Gaussian
distributions and mixing weights in Table 1 and Table 2 with
different maximum initial number of components. For the covariance
matrix, we use eigenvalues because the three components have the
same shape as $\mathcal{N}(\mbox{\boldmath{0}},\textrm{diag}(2,
0.2))$. Table 1 and Table 2 show that the modified EM algorithms
give accurate estimate for parameters and mixing weights. The final
estimate of these parameters is robust to the initialization of the
maximum number of components.

\begin{table}[htbp]
\begin{center} \small
\caption{Parameter Estimation (The initial number of components
$M=10$)} \label{T1}
\begin{tabular} {|c|c|c|c|c|c|c|} \hline
\multicolumn{2}{|c|}{Component} & Mixing Probability &  \multicolumn{2}{|c|}{Mean} &  \multicolumn{2}{|c|}{Covariance (eigenvalue)} \\ \hline
\multirow{3}{*}{1} & True &   0.3333 &  -1 & 1  &  2 & 0.2\\
& (\ref{e2.3}) & 0.3342(.0201) & -0.9911(.0861) & 1.0169(.1375) &
2.0034(.3022) &  0.1981(.0264) \\  & (\ref{e2.4}) & 0.3356(.0187) &
-1.0022 (.0875) & 1.0007(.1428) & 2.0205(.2769) & 0.1973 (.0265) \\
\hline
\multirow{3}{*}{2} & True & 0.3333 &  1 &1 &  2 & 0.2 \\
& (\ref{e2.3}) & 0.3317(.0196) & 1.0151(.0845) & 0.9849(.1318) &
1.9794(.2837) & 0.1977(.0303) \\
&(\ref{e2.4}) & 0.3321 (.0193) & 1.0108(.0790)& 0.9904(.1253)&
1.9825(.2980) & 0.1957 (.0292) \\ \hline
\multirow{3}{*}{3} & True & 0.3333 &  0 & -1.4142 &  2 & 0.2 \\
& (\ref{e2.3}) & 0.3341(.0171) & 0.0019(.1324) & -1.4112(.0405) &
1.9722(.2425) & 0.1973(.0258) \\
&(\ref{e2.4})& 0.3322(.0159)
&0.0014(.1449)&-1.4103(.0404)&1.9505(.2424) & 0.1978(.0267)
 \\  \hline
\end{tabular}
\end{center}
\end{table}

\begin{table}[htbp]
\begin{center} \small
\caption{Parameter Estimation (The initial number of components
$M=50$)} \label{T2}
\begin{tabular} {|c|c|c|c|c|c|c|} \hline
\multicolumn{2}{|c|}{Component} & Mixing Probability &
\multicolumn{2}{|c|}{Mean} &  \multicolumn{2}{|c|}{Covariance
(eigenvalue)} \\ \hline
\multirow{3}{*}{1} & True &  0.3333 &  -1 & 1  &  2 & 0.2\\
& (\ref{e2.3}) & 0.3342(.0201) & -1.0080(.0881) & 0.9854(.1372) &
1.9603(.2857) &  0.1974(.0296) \\  & (\ref{e2.4}) & 0.3320(.0190) &
-1.0017 (.0859) & 0.9985(.1389) & 1.9604(.2830) & 0.1960 (.0286) \\
\hline
\multirow{3}{*}{2} & True & 0.3333 &  1 &1 &  2 & 0.2 \\
& (\ref{e2.3}) & .3347(.0170) & 0.9879(.0885) & 1.0166(.1385) &
1.9531(.2701) & 0.1981(.0283) \\
&(\ref{e2.4}) & 0.3345 (.0182) & 0.9987(.0896)& 1.0044(.1402)&
1.9661(.2460) & 0.1971 (.0248) \\ \hline
\multirow{3}{*}{3} & True & 0.3333 &  0 & -1.4142 &  2 & 0.2 \\
& (\ref{e2.3}) & 0.3329(.0198) & 0.0210(.1329) & -1.4105(.0344) &
1.9717(.2505) & 0.1975(.0265) \\
&(\ref{e2.4})& 0.3334(.0164)
&0.0117(.1302)&-1.4116(.0372)&1.9736(.2769) & 0.1998(.0281)
 \\  \hline
\end{tabular}
\end{center}
\end{table}

\subsection{Example II}
In the second example, we consider a situation where the mixture
components overlap and may have same means but different covariance
matrices.  This is a rather challenging example, and the proposed
method by Chen and Khalili (2008) can not be applied as some
components have the same mean.  Specifically, we generate 1000
samples with mixing weights $\pi_1=\pi_2=\pi_3=0.3$, $\pi_4=0.1$,
mean vectors $\bmu_1=\bmu_2=\left[-2, -2\right]^T$, $\bmu_3=\left[2,
0\right]^T$, $\bmu_4 = \left[1, -4\right]^T$, and
\begin{eqnarray*}
\bSigma_1&=\left[\begin{array}{cc} 0.1 & 0 \\ 0 & 0.2 \end{array}\right],  \qquad\bSigma_2&=\left[\begin{array}{cc} 2 & 2 \\ 2 &  7\end{array}\right], \\
\bSigma_3&=\left[\begin{array}{cc} 0.5 & 0 \\ 0 & 4\end{array}\right], \qquad\bSigma_4&=\left[\begin{array}{cc} 0.125 & 0 \\ 0 & 0.125 \end{array}\right].
\end{eqnarray*}

\begin{figure}[htbp]
\c{\psfig{figure=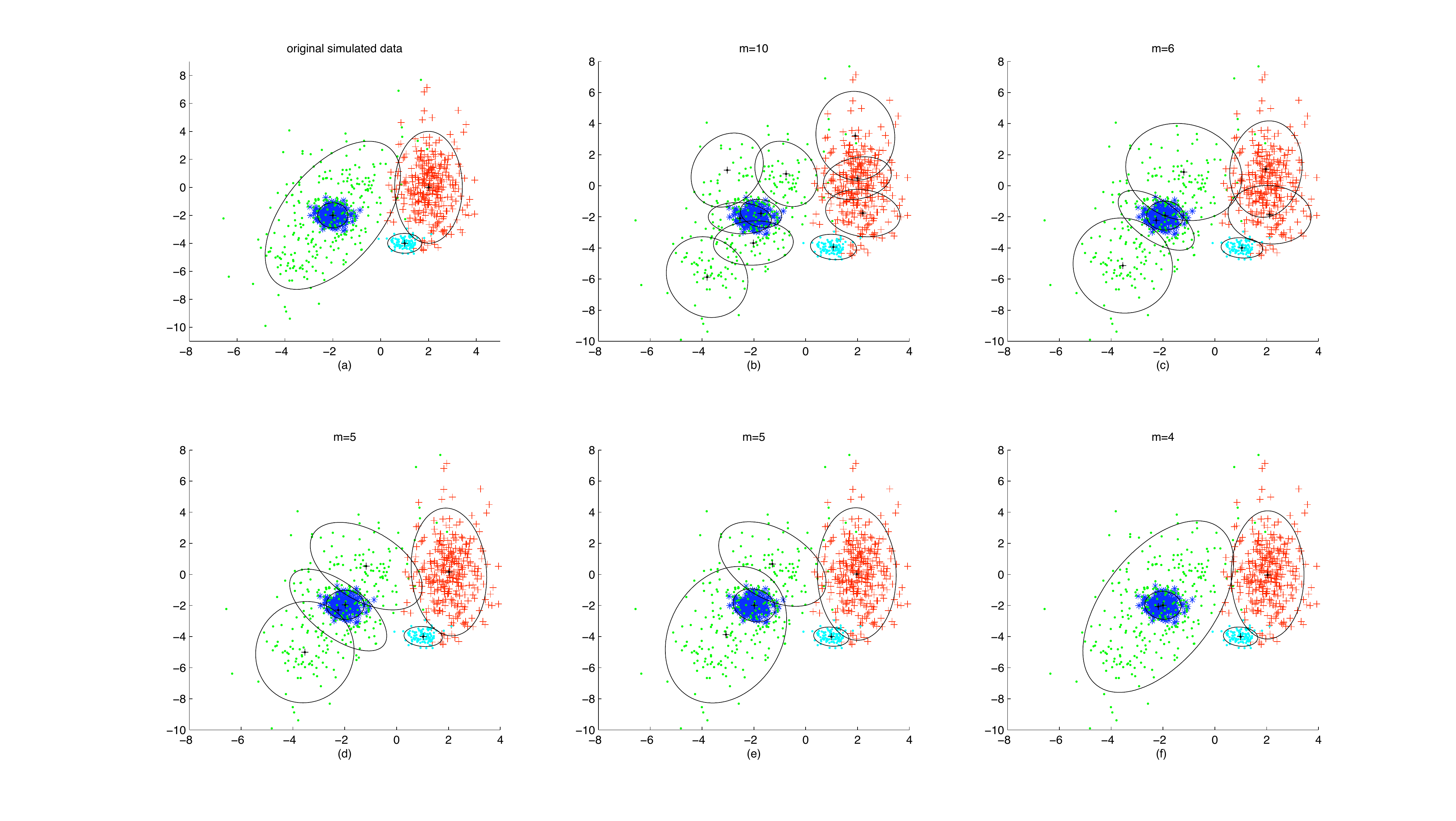,height=4in,width=6in}}
\begin{singlespace}
\caption{\em  \label{fig2s2}  One typical run.   (a) a simulated
data set. (b) initialization for $M=10$ components, (c-e) three
intermediate estimates for $M=7, 6, 5$, respectively, (f) the final
estimate for $M=4$. }
\end{singlespace}
\end{figure}

\begin{figure}[htbp]
\c{\psfig{figure=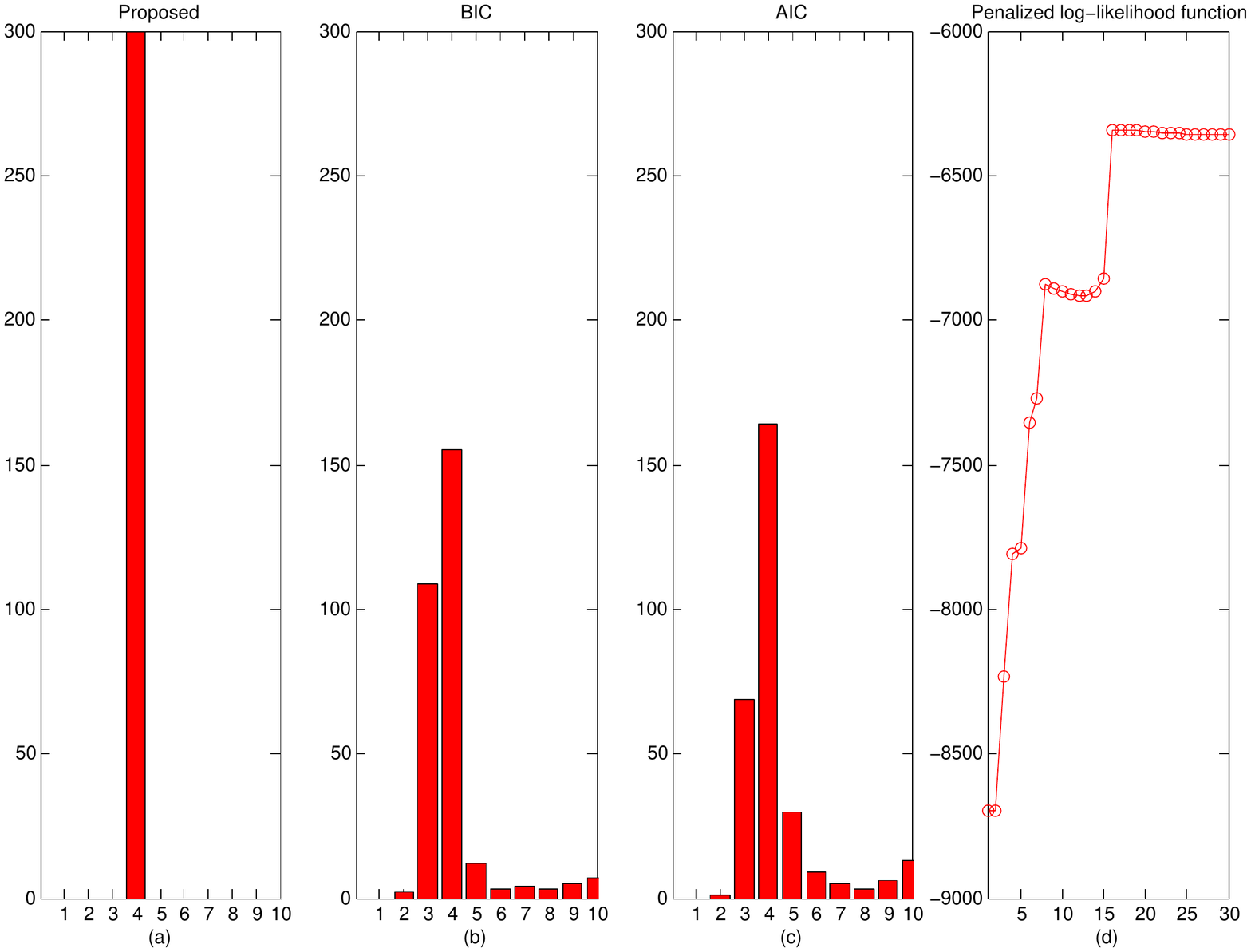,height=3in,width=6in}}
\begin{singlespace}
\caption{\em  \label{fig3s2} Histogram of  estimated numbers of
components. (a) the proposed method (2.3), (b) BIC, (c) AIC.  (d)
The penalized log likelihood function for one typical run. }
\end{singlespace}
\end{figure}

Similar to the first Example, we run our proposed methods for 300
times. The maximum number of components is set to be 10 or 50,  the
initial value for the modified EM algorithms is estimated by K-means
clustering, and the tuning parameter $\lambda$ is selected by our
proposed BIC method.  Figure~\ref{fig2s2} shows the evolution of the
modified EM algorithm for (2.3) with the maximum initial number of
components as 10 for one simulated data set. Figure~\ref{fig3s2}
shows that our proposed method can identify the number of components
$100\%$ correctly, and performs much better than AIC and BIC
methods. Table 3 and Table 4 show that the modified EM algorithms
give accurate estimates for parameters and mixing weights. Similar
as the first example, the final estimate of these parameters is
robust to the initialization of the maximum number of components.

\begin{table}[htbp] \label{T3} \small
\begin{center}
\caption{Parameter Estimation (The initial number of components
$M=10$)}
\begin{tabular} {|c|c|c|c|c|c|c|} \hline
\multicolumn{2}{|c|}{Component} & Mixing Probability & \multicolumn{2}{|c|}{Mean} &  \multicolumn{2}{|c|}{Covariance (eigenvalue)} \\ \hline
\multirow{3}{*}{1} & True &   0.3 &  -2 & -2  &  0.1 &  0.2\\
&(\ref{e2.3}) & 0.3022(.0093) & -2.0010(.0216) & -1.9989(.0291) &
0.0979(.0114) & 0.2010 (.0242) \\
& (\ref{e2.4}) & 0.3009(.0095) & -1.9995(.0206)& -1.9975(.0319) &
0.0990(.0119) & 0.2003(.0226) \\
\hline
\multirow{3}{*}{2} & True & 0.3 &  -2 & -2 &  1.2984 &  7.7016 \\
& (\ref{e2.3}) & 0.2995(.0112) & -1.9989(.1133)& -1.9963(.1837) &
1.2864(.1407)& 7.7219(.7301) \\
&(\ref{e2.4}) & 0.3017(.0118) & -1.9995(.1202)&
-2.0049(.1811)&1.2926(.1343)&7.5856(.7301) \\ \hline
\multirow{3}{*}{3} & True & 0.3 &    2 &  0 &  0.5 & 4 \\
& (\ref{e2.3}) & 0.3019(.0083) & 1.9943(.0483) & 0.0001(.1294)  &
0.4986(.0529)& 3.9951(.3496)
 \\
& (\ref{e2.4}) & 0.3012(.0087) & 1.9995(.0511) & -0.0001(.1244)  &
0.4963(.0544)& 3.9998(.3911) \\
  \hline
 \multirow{3}{*}{4} & True & 0.1 &   1 & -4 &  0.125 &  0.125 \\
& (\ref{e2.3}) & 0.0964(.0038) & 1.0005(.0373) &  -3.9966(.0394)  &
0.1143(.0245) & 0.1339(.0252)
 \\
& (\ref{e2.4}) & 0.0962(.0047) & 0.9993(.0394) &  -4.0013(.0417)  &
0.1167(.0259) & 0.1317(.0278) \\
  \hline
\end{tabular}
\end{center}
\end{table}

\begin{table}[htbp] \label{T4}
\begin{center} \small
\caption{Parameter Estimation (The initial number of components
$M=50$)}
\begin{tabular} {|c|c|c|c|c|c|c|} \hline
\multicolumn{2}{|c|}{Component} & Mixing Probability &
\multicolumn{2}{|c|}{Mean} &  \multicolumn{2}{|c|}{Covariance
(eigenvalue)} \\ \hline
\multirow{3}{*}{1} & True &   0.3 &  -2 & -2  &  0.1 &  0.2\\
&(\ref{e2.3}) & 0.3016(.0107) & -1.9982(.0223) & -1.9998(.0312) &
0.0986(.0110) & 0.2034 (.0241) \\
& (\ref{e2.4}) & 0.3009(.0095) & -1.9986(.0218)& -1.9978(.0320) &
0.0993(.0110) & 0.2010(.0238) \\
\hline
\multirow{3}{*}{2} & True & 0.3 &  -2 & -2 &  1.2984 &  7.7016 \\
& (\ref{e2.3}) & 0.3002(.0128) & -2.0040(.1086)& -2.0052(.1819) &
1.2823(.1386)& 7.6696(.7538) \\
&(\ref{e2.4}) & 0.3017(.0115) & -1.9988(.1173)&
-2.0116(.1811)&1.2757(.1318)&7.6734(.7476) \\ \hline
\multirow{3}{*}{3} & True & 0.3 &    2 &  0 &  0.5 & 4 \\
& (\ref{e2.3}) & 0.3015(.0083) & 1.9986(.0500) & 0.0054(.1365)  &
0.4998(.0531)& 3.9951(.3651)
 \\
& (\ref{e2.4}) & 0.3012(.0084) & 2.0015(.0505) & 0.0102(.1268)  &
0.4915(.0524)& 3.9751(.3770) \\
  \hline
 \multirow{3}{*}{4} & True & 0.1 &   1 & -4 &  0.125 &  0.125 \\
& (\ref{e2.3}) & 0.0966(.0044) & 0.9983(.0408) &  -4.0019(.0431)  &
0.1150(.0251) & 0.1327(.0258)
 \\
& (\ref{e2.4}) & 0.0962(.0050) & 1.0011(.0402) &  -4.0019(.0425)  &
0.1154(.0256) & 0.1313(.0254) \\
  \hline
\end{tabular}
\end{center}
\end{table}

\subsection{Real Data Analysis}
\begin{figure}[htbp]
\c{\psfig{figure=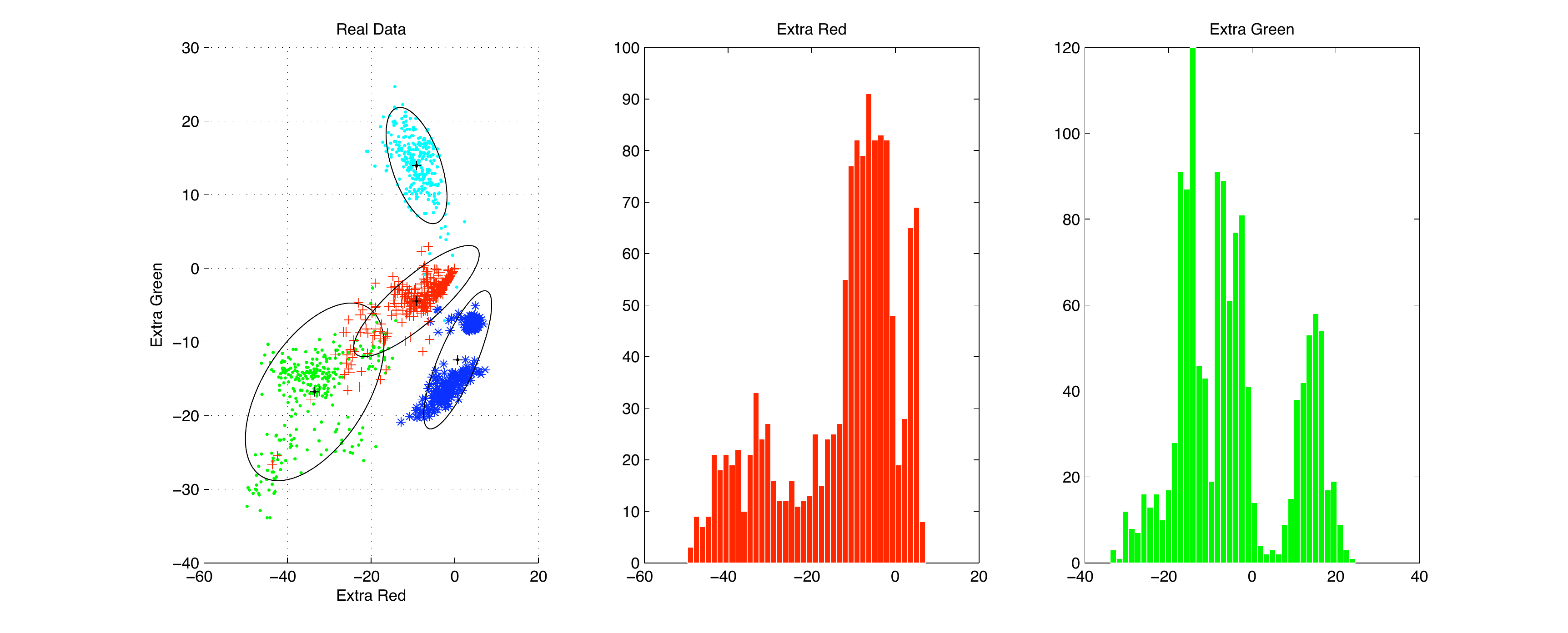,height=2in,width=6in}}  \hspace{1.5in}
(a)\hspace{1.5in} (b) \hspace{1.35in} (c)
\begin{singlespace}
\caption{\em  \label{figR1}  (a) Scatter plot of scaled real data.  Brickface (blue), Sky (green), Foliage (red), Grass (light blue); (b-c) Histograms of marginal density.  (b) Extra red, and (c) Extra green.  }
\end{singlespace}
\end{figure}

We apply our proposed methods  to an image segmentation data set at
UCI Machine Learning Repository
(http://archive.ics.uci.edu/ml/datasets/Image+Segmentation).   This
data set was created from a database of seven outdoor images
(brickface, sky, foliage, cement, window, path and grass). Each
image was hand-segmented into instances of $3\times3$ regions, and
230 instances were randomly drawn.  For each instance, there are 19
attributes.  We here only focus on four images, brickface, sky,
foliage and grass, and two attributes, extra red and extra green.
Our objective is to estimate the joint probability density function
of the two attributes (See Figure~\ref{figR1}(a)) using a Gaussian
mixture with arbitrary covariance matrices.  In other words, we
implement our proposed method to identify the number of components,
and to simultaneously estimate the unknown parameters of bivariate
normal distributions and mixing weights. Although we consider only
four images, Figure~\ref{figR1}(a) suggests that a five-component
Gaussian mixture is more appropriate and the brickface image is
better represented by two components.

\begin{figure}[htbp]
\c{\psfig{figure=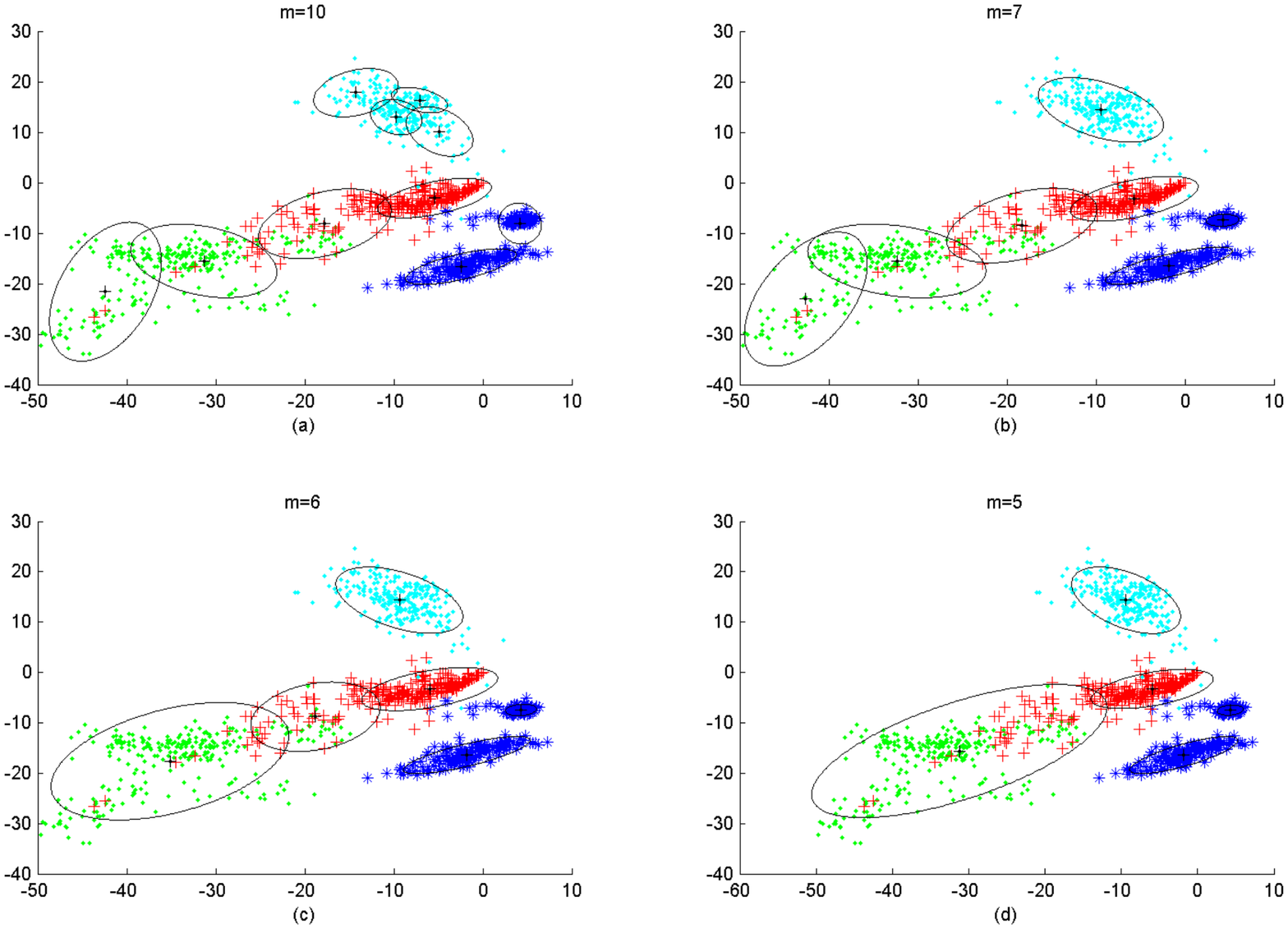,height=4in,width=6in}}
\begin{singlespace} \vspace{-0.35in}
\caption{\em  \label{figR2}  One typical run.  (a) Initialization with $M=10$ components, (b) and (c) two intermediate estimates for $M=7$ and $M=6$, respectively, (d) the final estimate $M=5$. }
\end{singlespace}
\end{figure}

\begin{figure}[htbp]
\c{\psfig{figure=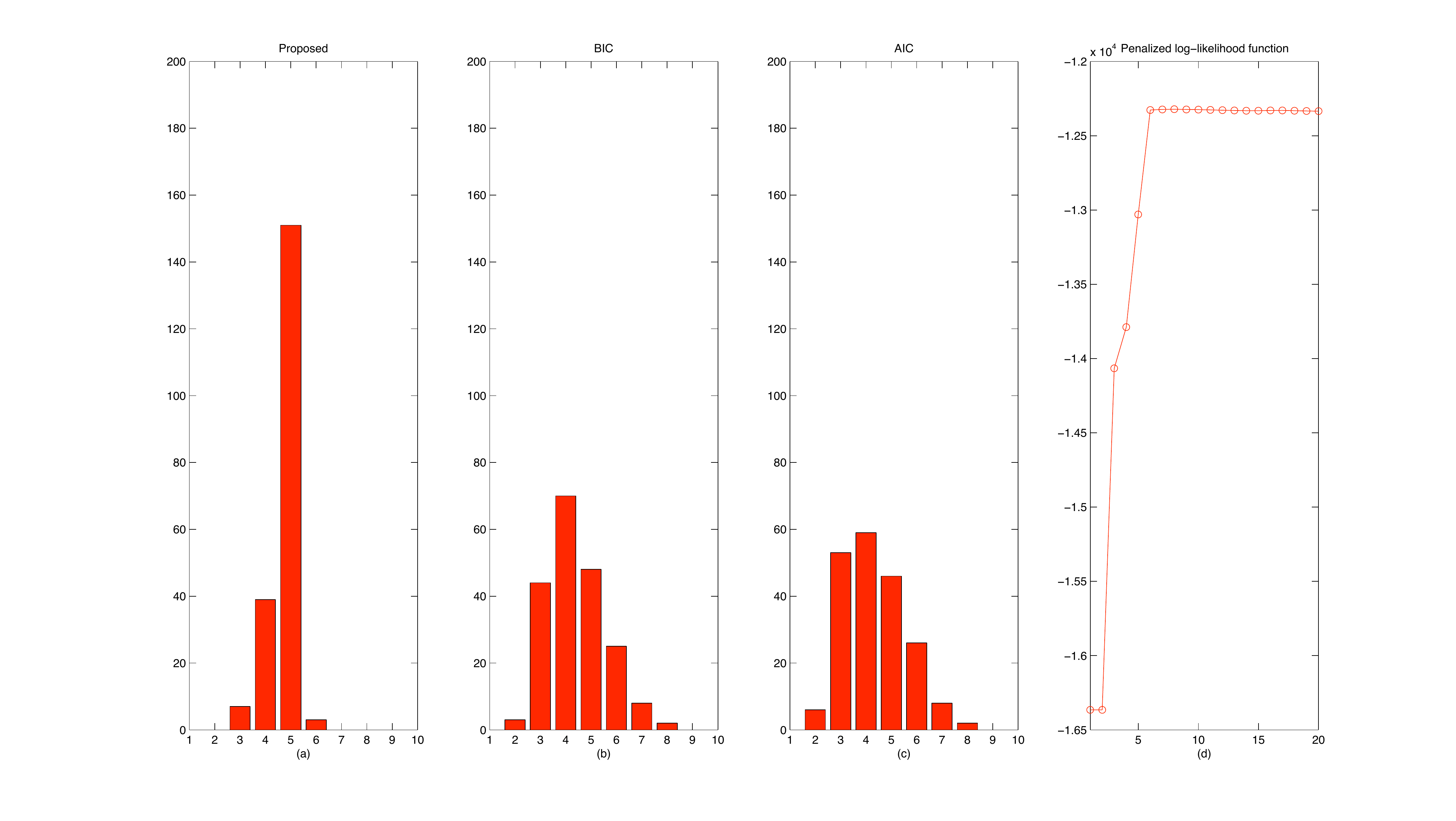,height=2in,width=6in}}
\begin{singlespace}
\caption{\em  \label{figR3} Histogram of  estimated numbers of
components. (a) the proposed method (2.3), (b) BIC, (c) AIC.  (d)
The penalized log likelihood function for one typical run.  }
\end{singlespace}
\end{figure}

As in the simulation studies, we run our proposed method for 300
times. For each run, we randomly draw 200 instances for each images.
The maximum number of components is set to be ten, and the initial
value for the modified EM algorithm is estimated by the K-means
clustering. Because there is little difference between numerical
results of the two proposed methods  (\ref{e2.3}) and (\ref{e2.4}),
here we only show the numerical results obtained by maximizing
(\ref{e2.3}). Figure \ref{figR2} shows the evolution of the modified
EM algorithm for one run. Figure \ref{figR3} shows that our proposed
method selects five components with high probability. For a
five-component Gaussian mixture model, we summarize the estimation
of parameters and mixing weights in Table 5.

\begin{table}[htbp]
\begin{center} {\tiny
\caption{Parameter Estimation, $\hat M= 5$} \label{T5}
\begin{tabular} {|c|c|c|c|c|c|c|} \hline
Component & Underlying &Mixing Probability & \multicolumn{2}{|c|}{Mean } &  \multicolumn{2}{|c|}{Standard Deviation }  \\
 & & & \multicolumn{2}{|c|}{(Ex-red, Ex-green)} &  \multicolumn{2}{|c|}{(Ex-red, Ex-green)}  \\ \hline
1  & Sky \& Foliage &  .4153(0.0555) & -27.7689(2.3336)  & -13.7343(1.1377)  &  12.0464(1.1726) &  7.3739(0.3745)  \\ \hline
2  & Grass  &.2617(0.0523) & -9.3724(0.8588)     & 13.4992(4.0740)   &  3.6393(1.2160)  & 3.5632(1.3567)  \\ \hline
3 &  Foliage \& Brickface & .1447(0.0425) & -4.9511(1.3913)     & -3.3625(3.5537)    &  3.1584(0.7102)    & 1.6728(0.4004)  \\ \hline
4 &  Brickface & .0824(0.0487) &  -1.3474(0.8417)      & -12.0906(7.2158)  & 2.5780(1.5227)   & 1.3302(0.7854) \\ \hline
5 & Brickface & .0936(0.0717)  & 3.5476(1.4703)    & -8.4996(2.1980)   &  1.5110(1.2288)   & 1.3293(1.6460) \\  \hline
 \end{tabular}
}
\end{center}
\end{table}

\section{Conclusions and Discussions}

In this paper, we propose a penalized likelihood approach for
multivariate finite Gaussian mixture models which integrates model
selection and parameter estimation in a unified way.  The proposed
method involves light computational load and is very attractive when
there are many possible candidate models.  Under mild conditions, we
have, both theoretically and numerically, shown that our proposed
method can select the number of components consistently for Gaussian
mixture models. Though we mainly focus on Gaussian mixture models,
we believe our method can be extended to more generalized mixture
models. This requires more rigorous mathematical derivations and
further theoretical justifications, and is beyond the scope of this
paper.

In practice, our proposed modified EM algorithm gradually discards
insignificant components, and does not generate new components or
split  any large components. If necessary, for very complex
problems, one can perform the split-and-merge operations (Ueda
\etal. 1999) after certain EM iterations to improve the final
results.  We only show the convergence of our proposed modified EM
algorithm through simulations, and further theoretical investigation
is needed.  Moreover, classical acceleration methods, such as Louis'
method, Quasi-Newton method and Hybrid method (McLachlan and Peel,
2000), may be used to improve the convergence rate of our proposed
modified EM algorithm.

Another practical issue is the selection of the tuning parameter
$\lambda$ for the penalized likelihood function. We propose a BIC
selection method, and simulation results show it works well.
Moreover, our simulation results show the final estimate is quite
robust to the initial number of components.


In this paper, we propose two penalized log-likelihood functions
(\ref{e2.3}) and (\ref{e2.4}). Although the numerical results
obtained by these two penalized functions are very similar, we
believe they have different theoretical properties. We have shown
the consistency of model selection and tuning parameter selection by
maximizing (\ref{e2.4}) and the BIC criterion proposed in the paper
under mild conditions.  We have also shown the consistency of model
selection by maximizing (\ref{e2.3}), but we note that the
conditions are somewhat restrictive. In particular, the consistency
of BIC criterion to select the tuning parameter for (\ref{e2.3})
needs further investigations.

An ongoing work is to investigate how to extend our proposed
penalized likelihood method to the mixture of factor analyzers
(Ghahramani and Hinton, 1997), and to integrate clustering and
dimensionality reduction in a unified way.

\section*{Appendix: Proof}
\renewcommand{\theequation} {A.\arabic{equation}}
\setcounter{equation}{0}

We now outline the key ideas of the proof for Theorem 3.1.

First assume the true Gaussian mixture density is
$$g_0=\sum\limits_{l=1}^q \pi_l^0\phi(\bmu_l^0,\bSigma_l^0),$$
then define $\mathcal{D}$ as the subset of functions of form
$$
\sum\limits_{l=1}^q \pi_l^0 \sum\limits_{i=1}^d
\frac{\delta_{\mu_i}^l D_i^1
\phi(\bmu^0_i,\bSigma^0_i)}{g_0}+\sum\limits_{l=1}^q \pi_l^0
\sum\limits_{i\ge j=1}^d \frac{\delta_{\Sigma_{i,j}}^l D_{i,j}^1
\phi(\bmu^0_l,\bSigma^0_l)}{g_0}+\sum\limits_{i=1}^{M-q} \lambda_i
\frac{\phi(\bmu_i,\bSigma_i)}{g_0}+\sum\limits_{l=1}^q \rho_l
\frac{\phi(\bmu_l^0,\bSigma_l^0)}{g_0}
$$
where $D_i^1$ is the derivative of $\phi(\mu_i,\bSigma_i)$ for the
$i$th component of $\mu_i$, $D_{i,j}^1$ is the derivative of
$\phi(\mu_i,\bSigma_i)$ for the $(i,j)$ component of $\bSigma_i$.
For functions in $\mathcal{D}$, $(\bmu_i,\bSigma_i), i=1,\ldots, M-q$
and $\bmu_l^0, \bSigma_l^0$ satisfy the conditions P1 and P2. The important
consequence for $\mathcal{D}$ is the following proposition. \\

\noindent {\bf Proposition A.1.} Under the conditions P1 and P2,
$\mathcal{D}$ is a Donsker class.

\noindent{\bf Proof:} First, by the conditions P1 and P2, it is easy to
check that the conditions P1 and P2 in Keribin (2000) or P0 and P1
in Dacunha-Castelle and Gassiat (1999) are satisfied by
$\mathcal{D}$. Then $( D_i^1 \phi(\bmu^0_i,\bSigma^0_i), D_{i,j}^1
\phi(\bmu^0_l,\bSigma^0_l), \phi(\bmu_i,\bSigma_i))$ are within
envelope functions $(F_1,F_2,F_3)$, and square integrable under
$g_0\nu$. On the other hand, by the restrictions imposed on the $\bbeta$,
$\lambda_i,\rho_l, \pi_l^0 \delta_{\mu_i}^l$ and $ \pi_l^0
\delta_{\Sigma_i}^l$ are bounded.  Therefore, similar to the proof of
Theorem 4.1 in Keribin (2000) or the proof of Proposition 3.1
in Dacunha-Castelle and Gassiat (1999), it is straightforward to show
that $\mathcal{D}$ has the Donsker property with the bracketing
number $N(\varepsilon)=1/\varepsilon^K$ where $K=M(d+d(d+1)/2)$. $\Box$ \\

\noindent{\bf Proof of Theorem 3.1:} To prove the theorem, we first
show that there exists a maximizer $(\theta, \bbeta)$ such that
$\theta=O_p(1/\sqrt{n})$. In fact, it is sufficient to show that,
for a large constant $C$, $\ell(\theta,\bbeta) < \ell(0,\bbeta)$
where $\theta=C/\sqrt{n}$. Let $ \theta=C/\sqrt{n}$, and notice that
\begin{eqnarray*}
  \ell_p(\theta,\bbeta)-\ell_p(0,\bbeta)
&=&  \sum\limits_{i=1}^n \{\log f(\bx_i,\theta,\bbeta)-\log
g_0(\bx_i)\}-n\lambda D_f\sum\limits_{m=1}^M [\log
(\epsilon+p_\lambda(\pi_m))-\log(\epsilon)] \\
& \ & + n\lambda D_f\sum\limits_{l=1}^q [\log
(\epsilon+p_\lambda(\pi^0_l))-\log(\epsilon)],
\end{eqnarray*}
and then
\begin{eqnarray*}
 \ell_p(\theta,\bbeta)-\ell_p(0,\bbeta)  & \le &
\sum\limits_{i=1}^n \{\log f(\bx_i,\theta,\bbeta)-\log g_0(\bx_i)\}
\\  & \ & -n\lambda D_f\sum\limits_{m=M-q+1}^M [\log
(\epsilon+p_\lambda(\pi_m))-
\log (\epsilon+p_\lambda(\pi^0_{m-M+q}))] \\
&\hat{=}& I_1+I_2.
\end{eqnarray*}
For $I_2$, because $ \theta=C/\sqrt{n}$ and by the restriction
condition on $\rho_l, l=1,\ldots, q$, we have
$|\pi_m-\pi_{m-M+q}^0|\le C/\sqrt{n}$ when $m>M-q$.  Due to the
property of the penalty function, we then have
\begin{eqnarray*}
|I_2| &=& |-n\lambda D_f\sum\limits_{m=M-q+1}^M [\log
(\epsilon+p_\lambda(\pi_m))- \log (\epsilon+p_\lambda(\pi^0_{m-M+q}))]| \\
&=& |-n\lambda D_f\sum\limits_{m=M-q+1}^M [\log (\epsilon+a\lambda)-
\log
(\epsilon+a\lambda)]| \\
&=& 0.
\end{eqnarray*}

For $I_1$, we have
\begin{eqnarray*}
I_1 &=& \sum\limits_{i=1}^n
\frac{f(\bx_i,\theta,\bbeta)-g_0(\bx_i)}{g_0(\bx_i)}-\frac12
\sum\limits_{i=1}^n \left(
\frac{f(\bx_i,\theta,\bbeta)-g_0(\bx_i)}{g_0(\bx_i)} \right)^2 +
\frac13 \sum\limits_{i=1}^n U_i\left(
\frac{f(\bx_i,\theta,\bbeta)-g_0(\bx_i)}{g_0(\bx_i)} \right)^3
\end{eqnarray*}
holds for $\theta=C/\sqrt{n}$, where $|U_i|\le 1$. Expand
$f(\bx,\theta,\bbeta)$ up to the second order,
$$
f(\bx,\theta,\bbeta)=g_0(\bx)+\theta\cdot f'(\bx, 0,\bbeta)
+\frac{\theta^2}{2}\cdot f''(\bx, \theta^\ast, \bbeta),
$$
for a $\theta^\ast \le \theta$.

Noticing $\theta=C/\sqrt{n}$, $\mathrm{E} f'/g_0 =0$ and $\mathrm{E}
f''/g_0 =0$, and by the conditions P1, P2 and Proposition A.1 for
the class $\mathcal{D}$,  we have
\begin{eqnarray*}
I_1=\left\{\sum\limits_{i=1}^n \theta
\frac{f'(\bx_i,0,\bbeta)}{g_0(\bx_i)}-\frac12\sum\limits_{i=1}^n
\theta^2
\left(\frac{f'(\bx_i,0,\bbeta)}{g_0(\bx_i)}\right)^2\right\}(1+o_p(1)).
\end{eqnarray*}
Since $ \frac{1}{\sqrt{n}}\sum\limits_{i=1}^n
\frac{f'(\bx_i,0,\bbeta)}{g_0(\bx_i)} $ converges uniformly in
distribution to a Gaussian process by Proposition A.1, and
$\sum\limits_{i=1}^n
\left(\frac{f'(\bx_i,0,\bbeta)}{g_0(\bx_i)}\right)^2$ is of order
$O_p(n)$ by the law of large numbers,  we have
$$
I_1=\frac{C}{\sqrt{n}}\cdot O_P(\sqrt{n})-\frac{C^2}{n}\cdot O_p(n).
$$
When $C$ is large enough, the second term of $I_1$ dominates other
terms in the penalized likelihood ratio. Then we have
$$  \ell_p(\theta,\bbeta)-\ell_p(0,\bbeta) <0$$
with probability tending to one.  Hence there exists a maximizer
$(\theta, \bbeta)$ with probability tending to one such that
$$
\theta=O_p(\frac{1}{\sqrt{n}}).
$$

\

Next we show that $\hat{q}=q$ or that $\hat{\pi}_m=0,
m=1,\ldots,M-q$, when the maximizer $(\theta, \bbeta)$ satisfies
$\theta=O_p(\frac{1}{\sqrt{n}})$. In fact, when
$\theta=O_p(\frac{1}{\sqrt{n}})$, we have
$\hat{\pi}_m=O_p(1/\sqrt{n}), m=1,\ldots, M-q,$ by the restriction
condition on $\lambda_i$. A Lagrange multiplier $\beta$ is taken
into account for the constraint $\sum_{m=1}^M \hat{\pi}_m=1$. Then
it is then sufficient to show that
\begin{equation}\label{A1}
\frac{\partial \ell^\ast(\btheta)}{\partial \hat{\pi}_m}<0 \quad
\mbox{for} \quad \hat{\pi}_m< \varepsilon_n
\end{equation}
with probability tending to one for the maximizer $(\theta, \bbeta)$
where $\varepsilon_n=Cn^{-1/2}$, $m\le M-q$, and $
\ell^\ast(\btheta)=\ell(\btheta)-\beta(\sum_{m=1}^M \pi_m-1).$ To
show the equation above, we consider  the partial derivatives for
$\hat{\pi}_m, \ m>M-q$ firstly. They should satisfy the following
equation,
\begin{equation}\label{A2}
\frac{\partial \ell^\ast(\btheta)}{\partial
\hat{\pi}_m}=\sum\limits_{i=1}^n
\frac{\phi_m(\mu_m,\bSigma_m)}{\sum_{i=1}^M \hat{\pi}_i
\phi_i(\mu_i,\bSigma_i)}-n\lambda D_f
\frac{1}{\epsilon+\hat{\pi}_m}-\beta=0.
\end{equation}
It is obvious that the first term in the equation above is of order
$O_p(n)$ by the law of large numbers.  If $m>M-q$ and
$\theta=O_p(\frac{1}{\sqrt{n}})$, it is easy to know that
$\hat{\pi}_m=\pi_{m-M+q}^0+O_p(1/\sqrt{n})>\frac12\cdot\min(\pi_1^0,\ldots,\pi_q^0)$,
and hence the second term should be $O_p(n\lambda)=o_p(n)$. So we
have $\beta=O_p(n)$.

Next, consider
\begin{equation} \label{A3} \frac{\partial
\ell^\ast(\btheta)}{\partial \hat{\pi}_m}=\sum\limits_{i=1}^n
\frac{\phi_m(\mu_m,\bSigma_m)}{\sum_{i=1}^M \hat{\pi}_i
\phi_i(\mu_i,\bSigma_i)}-n\lambda D_f
\frac{1}{\epsilon+\hat{\pi}_m}-\beta.
\end{equation}
where $m\le M-q$ and $\hat{\pi}_m<\varepsilon_n$. As shown for
(\ref{A3}), it is obvious that the first term and the third term
$\beta$ in the equation above are of order $O_p(n)$.  For the second
term, because $\pi_m=O_p(1/\sqrt{n})$, $\sqrt{n}\lambda \to \infty$
and $\epsilon$ is sufficient small, we have
$$
\left\{n\lambda D_f \frac{1}{(\epsilon+\pi_m)}\right\}\Big/n=\lambda
D_f \frac{1}{\epsilon+\pi_m}=O_p(\sqrt{n}\lambda) \to \infty.
$$
with probability tending to one. Hence the second term in the
equation (\ref{A3}) above dominates the first term and the third
term in the equation. Therefore we proved the equation (\ref{A1}),
or equivalently $\hat{\pi}_m=0,m=1,\ldots,M-q$ with probability
tending to one when $n \to \infty$. $\Box$ \\

\noindent{\bf Proof of Theorem 3.2:} To prove Theorem 3.2, similar
as the proof of Theorem 3.1,  we first show that there exists a
maximizer $(\theta, \bbeta)$ such that $\theta=O_p(1/\sqrt{n})$ when
$\lambda=C/\sqrt{n}$, and it is sufficient to show that, for a large
constant $C_1$, $\ell(\theta,\bbeta) < \ell(0,\bbeta)$ where
$\theta=C_1/\sqrt{n}$. Let $ \theta=C_1/\sqrt{n}$, and as the
similar step of Theorem 3.1, we have
\begin{eqnarray*}
  \ell_p(\theta,\bbeta)-\ell_p(0,\bbeta)
& \le & \sum\limits_{i=1}^n \{\log f(\bx_i,\theta,\bbeta)-\log
g_0(\bx_i)\} \\  & \ & -n\lambda D_f\sum\limits_{m=M-q+1}^M [\log
(\epsilon+\pi_m)-
\log (\epsilon+\pi_{m-M+q}^0] \\
&\hat{=}& I_1+I_2.
\end{eqnarray*}
For $I_2$, because of $ \theta=C_1/\sqrt{n}$ and by the restriction
condition on $\rho_l, l=1,\ldots, q$, we have
$|\pi_m-\pi_{m-M+q}^0|\le C_1/\sqrt{n}$ when $m>M-q$.  By the
property of the penalty function, we then have
\begin{eqnarray*}
|I_2| &=& |-n\lambda D_f\sum\limits_{m=M-q+1}^M [\log
(\epsilon+\pi_m)- \log (\epsilon+\pi^0_{m-M+q})]| \\
&=& \left|-n\lambda D_f\sum\limits_{m=M-q+1}^M \left[\frac{(\pi_m-\pi^0_{m-M+q})}{\epsilon+\pi^0_{m-M+q}}\cdot(1+o(1)) \right]\right| \\
&=& O(\sqrt{n})\cdot \frac{q C_1}{\sqrt{n}}(1+o(1))=O(C_1).
\end{eqnarray*}

For $I_1$, as in  the proof of Theorem 3.1, we have
$$
I_1=\frac{C_1}{\sqrt{n}}\cdot O_P(\sqrt{n})-\frac{C_1^2}{n}\cdot
O_p(n).
$$
When $C_1$ is large enough, the second term of $I_1$ dominates $I_2$
and other terms in $I_1$. Hence we have
$$  \ell_p(\theta,\bbeta)-\ell_p(0,\bbeta) <0$$
with probability tending to one.  Hence there exists a maximizer
$(\theta, \bbeta)$ with probability tending to one such that
$$
\theta=O_p(\frac{1}{\sqrt{n}}).
$$

\

Next we show that there exists a maximizer $(\hat{\theta},
\hat{\bbeta})$ satisfies $\hat{\theta}=O_p(\frac{1}{\sqrt{n}})$ such
that
 $\hat{q}=q$ or $\hat{\pi}_m=0, m=1,\ldots,M-q$,

First, we show that for any maximizer
$\ell_p(\theta^\ast,\bbeta^\ast)$ with $|\theta^\ast|\le
C_1/\sqrt{n}$ if there is $k \le M-q$ such that $C_1/\sqrt{n} \ge
\pi^\ast_k>1/\sqrt{n \log n}$, then there should exist another
maximizer of $\ell_p(\theta,\bbeta)$ in the area of $|\theta|\le
C_1/\sqrt{n}$. It means that the extreme maximizer of
$\ell_p(\theta,\bbeta)$ in the compact area $|\theta| \le
C_1/\sqrt{n}$ should satisfy that $\pi_k<\frac{1}{\sqrt{n}\log n}$
for any $k <M-q+1$. Hence it is also equivalent to show for any such
kind maximizer $\ell_p(\theta^\ast,\bbeta^\ast)$ with
$|\theta^\ast|\le C_1/\sqrt{n}$, we always have
$\ell_p(\theta^\ast,\bbeta^\ast)<\ell_p(0,\beta^\ast)$ with
probability tending to one. Similar as the analysis before, we have
\begin{eqnarray*}
  \ell_p(\theta^\ast,\bbeta^\ast)-\ell_p(0,\bbeta^\ast)
& \le & \sum\limits_{i=1}^n \{\log
f(\bx_i,\theta^\ast,\bbeta^\ast)-\log g_0(\bx_i)\} \\  & \ &
-n\lambda D_f\sum\limits_{m=M-q+1}^M [\log (\epsilon+\pi^\ast_m)-
\log (\epsilon+\pi_{m-M+q}^0]-n\lambda D_f\log\frac{\epsilon+\pi^\ast_k}{\epsilon} \\
&\hat{=}& I_1+I_2+I_3.
\end{eqnarray*}
As shown before, we have $I_1+I_2=O_p(C_1^2)$. For $I_3$, because
$\epsilon=o(\frac{1}{\sqrt{n}\log n})$ we have
$$
|I_3| = O(n \cdot {C/\sqrt{n}} )\cdot  \log
\frac{\pi^\ast_k}{\epsilon} =O ( \sqrt{n}).
$$
Then notice that $I_3$ is always negative and dominate $I_1$ and
$I_2$, and hence we have $\ell_p(\theta,\bbeta)<\ell_p(0,\beta)$.

In the following step, we need only consider the maximizer
$\ell_p(\hat{\theta},\hat{\bbeta})$ with $|\hat{\theta}|\le
C_1/\sqrt{n}$ and $\hat{\pi}_k<1/\sqrt{n \log n}$ for $k <M-q+1$.

A Lagrange multiplier $\beta$ is taken into account for the
constraint $\sum_{m=1}^M \hat{\pi}_m=1$.  It is then sufficient to
show that
\begin{equation}\label{A4}
\frac{\partial \ell^\ast(\btheta)}{\partial \hat{\pi}_m}<0 \quad
\mbox{for} \quad \hat{\pi}_m< \frac{1}{\sqrt{n}\log n}
\end{equation}
with probability tending to one for the maximizer $(\theta, \bbeta)$
where  $ \ell^\ast(\btheta)=\ell(\btheta)-\beta(\sum_{m=1}^M
\pi_m-1).$ To show the equation above, we consider  the partial
derivatives for $\hat{\pi}_m, \ m>M-q$ firstly. They should satisfy
the following equation,
\begin{equation}\label{A5}
\frac{\partial \ell^\ast(\btheta)}{\partial
\hat{\pi}_m}=\sum\limits_{i=1}^n
\frac{\phi_m(\mu_m,\bSigma_m)}{\sum_{i=1}^M \hat{\pi}_i
\phi_i(\mu_i,\bSigma_i)}-n\lambda D_f
\frac{1}{\epsilon+\hat{\pi}_m}-\beta=0.
\end{equation}
It is obvious that the first term in the equation above is of order
$O_p(n)$ by the law of large numbers.  If $m>M-q$ and
$\theta=O_p(\frac{1}{\sqrt{n}})$, it is easy to know that
$\hat{\pi}_m=\pi_{m-M+q}^0+O_p(1/\sqrt{n})>\frac12\cdot\min(\pi_1^0,\ldots,\pi_q^0)$,
and hence the second term should be $O_p(n\lambda)=o_p(n)$. So we
have $\beta=O_p(n)$.

Next, consider
\begin{equation} \label{A6} \frac{\partial
\ell^\ast(\btheta)}{\partial \hat{\pi}_m}=\sum\limits_{i=1}^n
\frac{\phi_m(\mu_m,\bSigma_m)}{\sum_{i=1}^M \hat{\pi}_i
\phi_i(\mu_i,\bSigma_i)}-n\lambda D_f
\frac{1}{\epsilon+\hat{\pi}_m}-\beta.
\end{equation}
where $m\le M-q$ and $\hat{\pi}_m<\frac{1}{\sqrt{n}\log n}$. As
shown for (\ref{A5}), it is obvious that the first term and the
third term $\beta$ in the equation above are of order $O_p(n)$.  For
the second term, because $\hat{\pi}_m=O_p(\frac{1}{\sqrt{n}\log
n})$, $\lambda=C/\sqrt{n}$ and $\epsilon =o(\frac{1}{\sqrt{n}\log
n})$ , we have
$$
\left\{n\lambda D_f \frac{1}{(\epsilon+\pi_m)}\right\}\Big/n=\lambda
D_f \frac{1}{\epsilon+\pi_m}=O_p(\lambda \cdot \sqrt{n} \log n ) \to
\infty.
$$
with probability tending to one. Hence the second term in the
equation (\ref{A6}) above dominates the first term and the third
term in the equation. Therefore we proved the equation (\ref{A4}),
or equivalently $\hat{\pi}_m=0,m=1,\ldots,M-q$ with probability
tending to one when $n \to \infty$. $\Box$

\

\noindent{\bf Proof of Theorem 3.3:} By Proposition A.1, and as in
the example of Gaussian Case shown in Keribin (2000), we know that
the Conditions (P1)-(P3) and (Id) are satisfied by Multivariate
Gaussian mixture model. Hence to prove this theorem, we can use the
theoretical results obtained by Keribin (2000) and follow the proof
step of Theorem 2 in Wang \etal. (2007).

First, given $\lambda^\ast=\sqrt{\frac{\log n}{n}}$, by Theorem 3.1,
we known that $\hat{q}=q$ with probability tending to 1, and
$\hat{\pi}_{M-q+m},m=1,\ldots q$ is the consistent estimate of
$\pi_i^0, i=1,\ldots, q$. Hence with probability tending to 1, we
have
$$\ell_P(\hat{\btheta}_{\lambda^\ast})=\ell(\hat{\btheta}_{\lambda^\ast})-n\lambda^\ast D_f \cdot q \cdot
\log \frac{\epsilon+a\lambda^\ast}{\epsilon}
$$
where $\btheta_{\lambda^\ast}$ is the parameter estimators of the
multivariate Gaussian mixture model. On the other hand, when $q$ is
known, we know its maximum likelihood estimate $\btheta_{MLE}$ is
consistent. Hence we have
\begin{eqnarray*}
\ell_P(\hat{\btheta}_{MLE}) &=&
\ell(\hat{\btheta}_{MLE})-n\lambda^\ast D_f \sum\limits_{m=1}^q
[\log (\epsilon+p_\lambda^\ast (\pi_{m, MLE}))-\log(\epsilon) ]  \\
&= & \ell(\hat{\btheta}_{MLE})-n\lambda^\ast D_f \cdot q \cdot \log
\frac{\epsilon+a\lambda^\ast}{\epsilon} \ge
\ell_P(\hat{\btheta}_{\lambda^\ast}),
\end{eqnarray*}
where $\pi_{m, MLE}$ is the maximum likelihood estimate of $\pi_m^0,
m=1,\ldots, q$.  Then by the convex property of $\ell(\btheta)$ and
the definition of $\ell_P(\hat{\btheta}_{\lambda^\ast})$,   when
$\lambda^\ast=\sqrt{\frac{\log n}{n}}$ we have the oracle property
of the penalized estimate of $\hat{\btheta}$ which should be equal
to $\hat{\btheta}_{MLE}$ with probability tending to one.

Next, we can identify two different cases, i.e., underfitting and
overfitting.

{\bf Case 1:} Underfitted model, i.e., $\hat{q}_\lambda<q$.
According to the definition of the BIC criterion, we have
$$
BIC_\lambda=\ell(\hat{\btheta}_\lambda)-\frac12 \hat{q}_\lambda D_f
\log n \le \ell(\hat{\btheta}_{\hat{q}_\lambda, MLE})-\frac12
\hat{q}_\lambda D_f \log n
$$
where $\hat{\btheta}_{\hat{q}, MLE}$ is the maximum likelihood
estimate of the finite Gaussian mixture model when the number of the
components is $\hat{q}_\lambda$. Similar as Keribin (2000), we know
that
$$
\frac{1}{n} \left\{\ell(\hat{\btheta}_{\hat{q}_\lambda,
MLE})-\ell(\hat{\btheta}_{q, MLE})\right\} =\frac{1}{n}
\left\{\ell(\hat{\btheta}_{\hat{q}_\lambda,
MLE})-\ell(\hat{\btheta}_{\lambda^\ast})\right\} \to -K(g_0,
\mathcal{G}_{\hat{q}_\lambda})
$$
where $\mathcal{G}_{\hat{q}_\lambda}$ is the finite Gaussian mixture
model space with $\hat{q}$ mixture components. Then we have
\begin{eqnarray*}
BIC_\lambda-BIC_{\lambda^\ast} &\le&
\ell(\hat{\btheta}_{\hat{q}_\lambda,
MLE})-\ell(\hat{\btheta}_{\hat{q}_{\lambda^\ast},
MLE})- \frac12 \hat{q}_\lambda D_f \log n+\frac12 \hat{q}_{\lambda^\ast} D_f \log n \\
&=& \ell(\hat{\btheta}_{\hat{q}_\lambda,
MLE})-\ell(\hat{\btheta}_{q, MLE})- \frac12 \hat{q}_\lambda D_f \log
n+\frac12 q D_f \log n \\
& = & -nK(g_0,
\mathcal{G}_{\hat{q}_\lambda}(1+o_p(1)) +\frac12(q-\hat{q}_\lambda) D_f \log n \\
& < & 0,
\end{eqnarray*}
This implies that
\begin{equation} \label{T31}
\mathrm{Pr}(\sup\limits_{\lambda:\hat{q}_\lambda<q
}BIC_\lambda>BIC_{\lambda^\ast}) \to 0.
\end{equation}

{\bf Case 2:} Overfitted model, i.e., $\hat{q}_\lambda>q$. As
Keribin (2000), for $p \ge q $, by Dacunha-Castelle (1999) we know
that
$$\ell(\hat{\btheta}_{\hat{q}_\lambda,
MLE})-\ell(\hat{\btheta}_{q, MLE})$$ converges in distribution to
the following variables
$$
\sup\left\{\sup\limits_{d \in \mathcal{D}} \frac12 \xi_d^2 1_{\xi_d
\ge 0} \ ; \sup\limits_{d_1 \in \mathcal{D}_1, d_2 \in
\mathcal{D}_2} \frac12 \left(\xi_{d_1}^2+\xi_{d_2}^2 1_{\xi_{d_2}
\ge 0}\right) \right\}
$$
where $\mathcal{D}_1$ and $\mathcal{D}_2$ are subsets of a unit
sphere $H$ of functions (For detail definition of $\mathcal{D}_1$,
$\mathcal{D}_1$ and $H$, see Keribin, 2000). Hence
$\ell(\hat{\btheta}_{\hat{q}_\lambda, MLE})-\ell(\hat{\btheta}_{q,
MLE})=O_p(1)$ and we have
\begin{eqnarray*}
BIC_\lambda-BIC_{\lambda^\ast} &\le&
\ell(\hat{\btheta}_{\hat{q}_\lambda,
MLE})-\ell(\hat{\btheta}_{\hat{q}_{\lambda^\ast},
MLE})- \frac12 \hat{q}_\lambda D_f \log n+\frac12 \hat{q}_{\lambda^\ast} D_f \log n \\
&=& \ell(\hat{\btheta}_{\hat{q}_\lambda,
MLE})-\ell(\hat{\btheta}_{q, MLE})- \frac12 \hat{q}_\lambda D_f \log
n+\frac12 q D_f \log n \\
& = & O_p(1) +\frac12(q-\hat{q}_\lambda) D_f \log n \\
& < & 0,
\end{eqnarray*}
and
\begin{equation} \label{T32}
\mathrm{Pr}(\sup\limits_{\lambda:\hat{q}_\lambda>q
}BIC_\lambda>BIC_{\lambda^\ast}) \to 0.
\end{equation}
Combined (\ref{T31}) with (\ref{T32}), Theorem 3.3 has been proved.
$\Box$

\vspace{0.5in}

\noindent{\large
 \bf References }
\begin{description}

\item{} Bishop, C. M. (2006).  Pattern Recognition and Machine Learning.
Springer.

\item{} Brand, M. E. (1999).  Structure learning in conditional probability models via an entropic prior and
parameter extinction. {\it Neural Computation}, {\bf
11}(5):1155-1182.

\item{} Chen, J. (1995).  Optimal rate of convergence for finite mixture models. {\it The Annals of Statistics}, {\bf 23}, 221-233.

\item{} Chen, J. and Kalbfleisch, J.D. (1996). Penalized minimum-distance estimates in finite mixture models. {\it Canadian Journal of Statistics}, {\bf 24}, 167-175.

\item{} Chen, J. and Khalili A. (2008).  Order selection in finite mixture models with a nonsmooth penalty. {\it
Journal of the American Statistical Association}, {\bf 104},    187-196.

\item{} Corduneanu, A.  and Bishop, C. M. (2001).  Variational Bayesian Model Selection for
Mixture Distributions. In {\it Proceedings Eighth International
Conference on Artificial Intelligence and Statistics}, 27--34,
Morgan Kaufmann.

\item{} Dacunha-Castelle, D. and Gassiat, E. (1997). Testing in
locally conic models and application to mixture models. {\it ESAIM:
P\& S (Probability and Statistics)} 285-317.\\
http://www.edpsciencs.com/ps/.

\item{} Dacunha-Castelle, D. and Gassiat, E. (1999). Testing the
order of a model using locally  conic parametrization: population
mixtures and stationary ARMA processes, {\it The Annals of
Statistics}, {\bf 27}, 1178-1209.

\item{} Dempster, A.P., Laird, N.M. and Rubin, D.B. (1977). Maximum likelihood from incomplete
data via the EM algorithm (with discussion), {\it Journal of the Royal Statistical Society, Ser. B},  {\bf 39},
1-38.

\item{} Fan, J. and Li, R. (2001). Variable selection via
nonconcave penalized likelihood and its oracle properties, {\it
Journal of the American Statistical Association}, {\bf 96},
1348-1360.

\item{} Figueiredo, M. and Jain, A.(2002). Unsupervised learning on
finite mixture models, {\it IEEE Transaction on Pattern analysis and
Machine intelligence} {\bf 24}, 381-396.

\item{} Ghahramani, Z. and Hinton, G.E. (1997).  The EM algorithm for mixtures of factor analyzers. Technical Report CRG-TR-96-1, University of Toronto, Canada.

\item{} James, L.F., Priebe, C.E., and Marchette, D.J. (2001).  Consistent estimation of mixture complexity, {\it The Annals of Statistics}, {\bf 29}, 1281-1296.

\item{} Keribin, C. (2000). Consistent estimation of the order of
mixture models. {\it Sankhy\={a}}, A,  {\bf 62}, 49-66.

\item{} Leroux, B. (1992). Consistent estimation of a mixing distribution. {\it The Annals of of Statistics}, {\bf 20}, 1350-1360.

\item{} Lindsay, B.G. (1995).  {\it Mixture Models: Theory, Geometry and Applications}. NSF-CBMS Regional Conference Series in Probability and Statistics, Volume 5, Institute for Mathematical Statistics: Hayward, CA.

\item{} McLachlan, G. and Peel, D. (2000). {\it Finite Mixture Models}. John Wiley \& Sons, New York.

\item{} Ormoneit, D. and Tresp, V. (1998).  Averaging, maximum penalized likelihood and Bayesian
estimation for improving Gaussian mixture probability density
estimates. {\it IEEE Transactions on Neural networks}, {\bf 9}(4):
1045--9227.

\item{} Ray, S. and Lindsay, B.G. (2008). Model selection in High-Dimensions: A Quadratic-risk Based Approach
.  {\it Journal of the Royal Statistical Society, Series B}, 70(1),
95-118.

\item{} Roeder, K. and Wasserman, L. (1997). Practical density
estimation using mixtures of normals. {\it Journal of the American
Statistical Association} {\bf 92}, 894-902.

\item{} Tibshirani, R. J. (1996). Regression Shrinkage and
Selection via the LASSO, {\it Journal of the Royal Statistical
Society, Ser. B}, {\bf 58}, 267-288.

\item{} Ueda, N. and Nakano, R. (1998).  Deterministic annealing EM algorithm.  {\it Neural Networks}, {\bf 11}, 271-282.

\item{} Ueda, N., Nakano, R., Ghahramani, Z., and Hinton, G.E. (1999).  SMEM algorithm for mixture models. {\it Advances in Neural Information
Processing Systems}, {\bf 11}, 299-605.

\item{} Wang, H., Li, R. and Tsai, C.-L. (2007). Tuning parameter selectors for the
smoothly clipped absolute deviation method. {\it Biometrika}. {\bf
94}, 553-568.

\item{} Woo, M., and Sriram, T.N. (2006).  Robust estimation of mixture complexity. {\it Journal of the American Statistical Association}, {\bf 101}, 1475-1485.

\item{} Zivkovic, Z. and van der Heijden, F. (2004). Recursive unsupervised
learning of finite mixture models. {\it IEEE Transactions on Pattern
Analysis and Machine Intelligence},  {\bf 26}(5), 651-- 265.

\end{description}

\end{document}